# Can sunlight replace LEDs in vertical farms? A critical assessment of optical-fiber daylighting strategies

Francesco Ceccanti[1]*, Aldo Bischi[1], Marco Antonelli[1], Andrea Baccioli[1]

1 Department of Energy, Systems, Territory and Construction Engineering
Università di Pisa, Pisa, Italy

*Corresponding Author: francesco.ceccanti@phd.unipi.it (F. Ceccanti)

**Abstract**

Vertical farming can increase food production while reducing land and water use, but its high electricity demand, mainly driven by artificial lighting, remains a major limitation. This simulation-based study evaluates the techno-economic potential of an optical-fiber daylighting system for a vertical farm located in Dubai. A previously validated transient vertical-farm model developed by the authors was coupled with Tonatiuh ray-tracing simulations to assess a single-axis solar-tracked Fresnel concentration system. Five optical-fiber typologies and three daylighting strategies were compared in terms of crop production, specific electrical energy consumption, and light cost, and benchmarked against light-pipe and rooftop photovoltaic solutions. The optical-fiber system achieved an annual useful-PAR delivery efficiency of 19–27%. By distributing daylight across all rack shelves and limiting heat gains through a UV–IR filter, it reduced electricity consumption by 35–89%, compared with 12% for the light-pipe configuration. The daylight-only strategy achieved the lowest SEEC, 1.81 kWh $kg^{-1}$, but reduced crop productivity by up to 52%. Hybrid LED–optical-fiber strategies largely avoided this penalty, achieving electricity savings of about 50% and a minimum SEEC of 3.16 kWh $kg^{-1}$. However, current fiber costs remain prohibitive, with an estimated CAPEX reduction of about 90% required for competitiveness. Rooftop PV covered 22% of annual electricity demand with an 11–12-year payback. Overall, sunlight can substantially reduce LED use but cannot currently replace it without significant agronomic or economic penalties.

**Keywords:**
Controlled environment agriculture; Vertical farming; Daylighting; Solar Energy; Optical Fibers; Energy System; Energy Consumption

Nomenclature

| **Acronyms** | | | **Subscripts** | |
|---|---|---|---|---|
| **AHU** | Air Handling Unit | $[-]$ | **abs** | Absorbed |
| **CAPEX** | Capital Expenditures | $[\$]$ | **ahu** | Air Handling Unit |
| **CEA** | Controlled Environment Agriculture | $[-]$ | **BOS** | Balance of System |
| **COP** | Coefficient of Performance | $[kW_{th}\ kW_{el}^{-1}]$ | **cond** | Conduction |
| **DLI** | Daily Light Integral | $[\mu mol\ m^{-2}\ day^{-1}]$ | **conv** | Convection |
| **HVAC** | Heating, Ventilation and Air Conditioning | $[-]$ | **$CO_2$** | Carbon Dioxide |
| **LAI** | Leaf Area Index | $[m_{leaf}{}^{2}\ m_{soil}{}^{-2}]$ | **diff** | Diffuse |
| **LC** | Light Cost | $[\$\ (\mu mol\ s^{-1})^{-1}]$ | **dir** | Direct |
| **LP** | Light Pipe | $[-]$ | **dl** | Daylight |
| **LUE** | Light Use Efficiency | $[g\ \mu mol^{-1}]$ | **dm** | Dry Matter |
| **PAR** | Photosynthetically Active Radiation | $[\mu mol\ m^{-2}\ s^{-1}]$ | **env** | Envelope |
| **PBT** | Payback Time | $[years]$ | **eva** | Evaporation |
| **POF** | Plastic Optical Fiber | $[-]$ | **fm** | Fresh Matter |
| **PPE** | Photosynthetic Photon Efficiency | $[\mu mol\ J^{-1}]$ | **H/C** | Heating & Cooling Systems |
| **PPF** | Photosynthetic Photon Flux | $[\mu mol\ s^{-1}]$ | **hum** | Humidification System |
| **PPFD** | Photosynthetic Photon Flux Density | $[\mu mol\ m^{-2}\ s^{-1}]$ | **In** | Inlet |
| **PV** | Photovoltaic | $[-]$ | **inv** | Inverter |
| **SEC** | Specific Energy Consumption | $[kWh\ kg^{-1}]$ | **LED** | Lighting System |
| **SEEC** | Specific Electric Energy Consumption | $[kWh\ kg^{-1}]$ | **OF** | Optical Fiber |
| **SLA** | Specific Leaf Area | $[cm^2\ g^{-1}]$ | **plant** | Single Lettuce Plant |
| **SOF** | Silica Optical Fiber | $[-]$ | **rad** | Radiative |
| **VF** | Vertical Farm | $[-]$ | **sol** | Solar |
| **Symbols** | | | **T** | Temperature |
| $\boldsymbol{A}$ | Area | $[m^2]$ | **tr** | Transmitted |
| $\boldsymbol{c_p}$ | Specific Heat Capacity | $[J\ kg^{-1}\ K^{-1}]$ | | |
| $\boldsymbol{d}$ | Density | $[kg\ m^3]$ | | |
| $\boldsymbol{DM}$ | Dry Matter | $[g\ m^{-2}]$ | | |
| $\boldsymbol{f}$ | Focal Length | $[mm]$ | | |
| $\boldsymbol{FM}$ | Fresh Matter | $[-]$ | | |
| $\boldsymbol{I}$ | Solar Irradiation | $[-]$ | | |
| $\boldsymbol{k}$ | Light Extinction Coefficient | $[-]$ | | |
| $\boldsymbol{n}$ | Index of Refraction | $[-]$ | | |
| **NA** | Numerical Aperture | $[-]$ | | |
| $\boldsymbol{Q}$ | Thermal Power | $[kW]$ | | |
| $\boldsymbol{P}$ | Electrical Power | $[kW]$ | | |
| $\boldsymbol{T}$ | Temperature | $[°C]$ | | |
| $\boldsymbol{t}$ | Time | $[s]$ | | |
| $\boldsymbol{V}$ | Volume | $[m^3]$ | | |
| $\boldsymbol{W}$ | Width | $[mm]$ | | |
| $\boldsymbol{Z}$ | Receiver Axial Displacement | $[mm]$ | | |
| **Greeks** | | | | |
| $\boldsymbol{\alpha}$ | Altitude | $[rad]$ | | |
| $\boldsymbol{\gamma}$ | Azimuth Angle | $[rad]$ | | |
| $\boldsymbol{\eta}$ | Efficiency | $[-]$ | | |
| $\boldsymbol{\tau}$ | Optical Transmission | $[-]$ | | |
| $\boldsymbol{\theta}$ | Incidence Angle | $[rad]$ | | |
| $\nu$ | Semi-angle of Fresnel Lens | $[rad]$ | | |
| $\rho$ | Optical Reflectance | $[-]$ | | |

## 1. Introduction

Vertical farming is a controlled crop production approach in which plants are grown in vertically stacked layers within highly managed indoor or semi-indoor environments. It is generally considered part of Controlled Environment Agriculture (CEA), a broader group of systems designed to regulate crop growth conditions independently from outdoor climatic variability [1]. By increasing the cultivated surface per unit of land footprint, vertical farms can enhance spatial productivity and are therefore frequently discussed as a potential strategy for food production in urban areas, where agricultural land is scarce or economically constrained [2]. These systems commonly rely on soilless cultivation methods, such as hydroponics or aeroponics, which allow a precise supply of water and nutrients and can reduce the dependence on soil availability and conventional irrigation practices [3]. The high level of environmental control also enables year-round production, improved protection from external weather conditions, and reduced exposure to pests and diseases, making vertical farming particularly suitable for compact, high-value crops such as leafy vegetables, herbs, and other short-cycle horticultural species [4]. For these reasons, vertical farming has been proposed as a complementary production model to conventional agriculture, with its potential benefits depending on crop type, local resource availability, climate, and the overall efficiency of the production system [5].

Vertical farming offers several potential advantages in terms of land productivity, resource-use efficiency, and production stability. By stacking multiple cultivation layers within the same footprint, these systems can substantially increase land-use efficiency, making crop production feasible in dense urban areas or regions where arable land is limited [5]. For leafy vegetables, recent reviews report lettuce yields in vertical farms of approximately 60–105 kg fresh weight $m^{-2}$ $year^{-1}$, depending on crop type, system design, and management strategy [6] [7]. Closed-loop soilless cultivation can also improve water-use efficiency by reducing runoff, drainage losses, and soil evaporation; for lettuce, an average water-use efficiency of about 140 g fresh weight $L^{-1}$ $H_2O$ has been reported [6]. In addition, the controlled indoor environment enables year-round production, reduces exposure to extreme weather, pests, and diseases, and can support food production closer to consumers, contributing to shorter and more resilient supply chains [8].

Despite these potential benefits, vertical farming still faces major limitations related to energy demand, economic viability, and scalability. Since indoor production replaces several functions normally provided by the outdoor environment, vertical farms require continuous energy inputs for lighting, climate control, air circulation, water pumping, and automation. Recent estimates indicate that operational vertical farms may convert purchased electricity into edible plant energy with an overall efficiency of only 1–2%, while a commercial lettuce case reported about 5 million kWh $year^{-1}$ for 500,000 kg of lettuce, corresponding to roughly 10 $kWh_{el}$ $kg^{-1}$ of fresh product [9]. Other reviews report energy demands of 10–18 kWh $kg^{-1}$ for lettuce production in stacked systems, compared with less than 1 kWh $kg^{-1}$ in field production [10]. Techno-economic analyses further indicate that the economic performance of vertical farming is strongly influenced by both local electricity tariffs and site-specific climatic conditions. An earlier study from the present research group on lettuce production in growth chambers reported substantial differences in annual energy requirements across different cities, with energy-related costs ranging from approximately 0.46 to 1.74 USD $kg^{-1}$ of fresh product [11]. These constraints are further amplified by high upfront investment, technical complexity, limited operational data, and the strong coupling between lighting, HVAC, and dehumidification loads, making energy reduction and low-carbon electricity supply central requirements for improving the sustainability of vertical farming [12].

Artificial lighting is one of the most energy-intensive functions in fully indoor vertical farming, because the photosynthetically active radiation (PAR) required for plant growth must be generated electrically. Lighting is generally provided by LED fixtures installed close to the crop canopy and operated according to predefined crop-specific light recipes, which combine photosynthetic photon flux density (PPFD), photoperiod, spectral composition, and daily light integral (DLI) [13]. These parameters are selected according to crop species and growth stage, and can be implemented through fixed schedules, dimming strategies, zonal control, or sensor-based systems that adjust light output to maintain target photon delivery [14]. Although LEDs offer high photon efficacy, tunable spectra, and precise controllability, they also introduce large internal heat gains and influence crop transpiration, thereby coupling lighting operation with cooling and dehumidification requirements. This coupling is reflected in reported electricity breakdowns: in a tested container farm, LED lighting accounted for 72% of daily electricity use, while HVAC and auxiliary loads represented 14% each [15]. Similar trends were reported in earlier work from the present research group, where artificial lighting was identified as the dominant electrical load in vertical farming systems [16]. For example, in an agrivoltaic agrotunnel case study, the annual electricity demand for lighting was estimated at 58,342 kWh, far exceeding the demand of the heat pump, dehumidifier, and pumps [17]. A comparison between plant factories and greenhouses further highlights the energetic penalty associated with replacing solar radiation with electric light. For lettuce production, the plant factory required 247 kWhe $kg^{-1}$ of dry biomass, compared with 70 kWhe $kg^{-1}$ in a greenhouse located in the Netherlands, where solar radiation supplied most of the photosynthetic energy at no direct electricity cost. Although the plant factory achieved higher resource-use efficiency and lower total specific energy input, artificial lighting accounted for approximately 90% of its electrical load [2]. These findings indicate that reducing the electricity required for photon delivery, while managing the associated HVAC impacts, is a central challenge for improving the energy performance of vertical farming systems.

To reduce the electricity demand associated with artificial photon delivery, a growing branch of research has investigated the integration of daylighting strategies into controlled indoor and vertical farming systems. Daylighting can be broadly defined as the collection, transmission, distribution, and control of natural solar radiation within an indoor space, with the aim of partially replacing or complementing electric lighting [18]. In building applications, daylighting systems have traditionally been developed to reduce lighting electricity demand and improve indoor environmental quality. When applied to plant production, however, their objective shifts from visual illumination to the delivery of PAR to the crop canopy. Unlike conventional greenhouses, where sunlight enters directly through transparent envelopes, daylighting strategies for vertical farming aim to introduce natural light into enclosed or multilayer systems while preserving a high degree of environmental control. These strategies may include transparent or semi-transparent envelopes, solar tubes, mirrors, concentrating optics, light guides, and optical fiber systems. Their potential advantage lies in the possibility of using solar radiation directly, avoiding part of the conversion chain from sunlight to electricity and then to artificial light. Nevertheless, their application in vertical farming requires careful control of spatial light distribution, spectral quality, seasonal variability, and unwanted thermal gains, particularly those associated with infrared radiation. For this reason, daylighting in vertical farming is generally explored as a hybrid strategy, in which natural light is used to offset part of the artificial lighting demand, while LEDs remain necessary to guarantee stable PPFD, photoperiod, and DLI targets under variable outdoor solar conditions [14].

Several studies have investigated the integration of daylighting strategies into vertical farming systems. The first group of approaches exploits direct or façade-mediated solar access, but this strategy is strongly limited by external shading and by self-shading among stacked cultivation layers. Eaton et al., for example, modelled natural light availability in skyscraper farms and showed that surrounding buildings can drastically reduce the amount of useful solar radiation reaching crops, in some cases allowing daylight to cover only about 5% of the total crop light requirement [19]. To improve daylight penetration in multi-tier systems, Lee et al. investigated sunlight-redirecting panels and reported an increase in photosynthetic activity of approximately 10–12%, showing that optical redirection can partially mitigate the non-uniform distribution of natural light across vertical racks [20]. Similarly, An et al. proposed a hybrid louver lighting system combining façade-mounted reflectors with infrared-cutting glazing, achieving a 33.8% reduction in electricity consumption under sunny-day conditions compared with a fully LED-lit reference vertical farm [21]. Other studies have focused on hybrid natural–artificial lighting control. Pereira and Gomes developed a sensor-based LED control strategy in a vertical urban farm greenhouse, where artificial light was supplied after sunset to meet predefined DLI targets of 12.5, 15, and 17.5 mol $m^{-2}$ $d^{-1}$. In that study, LED electricity use ranged from 1.64 to 4.90 kWh $kg^{-1}$, with the lowest DLI treatment requiring artificial lighting for only 16% of the total light supply [14]. An earlier light-pipe study by the same authors assessed roof-integrated light pipes in a three-tier container vertical farm in Dubai. Their ray-tracing simulations estimated a crop-level optical efficiency of 45–75%, while year-round dynamic simulations showed that daylight-only operation reduced electricity use by 27–29% but caused a 17% yield penalty [22]. Hybrid daylight–LED strategies preserved crop yield while reducing specific electric energy consumption, with PWM dimming combined with UV–IR filtering achieving 6.32 kWh $kg^{-1}$, about 14% lower than the fully LED benchmark [22]. Overall, these results suggest that daylighting can reduce the electrical burden of artificial lighting in vertical farming, but only when natural light is delivered with sufficient spatial uniformity, filtered or controlled to limit non-productive heat gains, and integrated with supplemental LEDs to preserve stable PPFD and DLI targets.

A more targeted approach to daylighting in vertical farming is the use of optical fiber daylighting systems, in which sunlight is collected outdoors, concentrated, transported through optical fibers, and redistributed close to the crop canopy. Compared with façade- or glazing-based daylighting, optical fibers can potentially deliver natural light directly to specific growing layers while limiting the penetration of unwanted ultraviolet and infrared radiation [23] [24]. Early work by Asiabanpour et al. investigated PMMA optical fibers for indoor plant growth through simulations and empirical tests, focusing on the effects of fiber diameter and distance from the emitter on light intensity and uniformity. Their results showed that direct sunlight reached 2121 μmol $m^{-2}$ $s^{-1}$, whereas the PPFD delivered through a 20 mm optical fiber was 69.3 μmol $m^{-2}$ $s^{-1}$ without a UV filter and 15.1 μmol $m^{-2}$ $s^{-1}$ with a UV filter, indicating reductions of approximately 30 and 140 times compared with direct sunlight, respectively [23]. Vu et al. subsequently developed and tested an optimized optical fiber daylighting system for indoor agriculture based on a Fresnel lens array and plastic optical fiber bundles. The prototype achieved a geometric concentration ratio of 140.4, an optical efficiency of approximately 19% in simulation and 16% experimentally. When installed in an indoor farm, the system reduced artificial lighting energy consumption by 54.3% on a sunny day and 38.9% on a cloudy day compared with LED-only lighting [24]. Other studies have combined fiber-guided solar illumination with spectral-management strategies. Yalçın and Ertürk proposed fluorescent reflectors to improve light distribution and convert less photosynthetically effective wavelengths into more useful spectral bands, estimating crop production increases exceeding 35% under selected configurations [25]. Building on this concept, Kaya et al. analyzed solar-illuminated vertical farms using fiber-optic light distribution, infrared filters, fluorescent coatings, and supplemental LEDs across three latitudes. Their results showed that hybrid solar illumination strategies with IR filtering, fluorescent coatings, supplemental LED lighting, and ventilation could reduce energy use by 25% relative to LED-only illumination [26]. Taken together, these studies suggest that optical fibers can reduce the reliance on electric lighting by transporting useful solar radiation into controlled growing environments. However, the reported performance remains strongly dependent on optical efficiency, collection area, spectral filtering, distribution uniformity, tracking complexity, and the additional thermal loads introduced by solar gains.

Although daylighting strategies have increasingly been investigated as a means to reduce the electricity demand associated with artificial lighting in vertical farming, the existing literature still presents two main limitations. First, most studies focus on optical performance, lighting energy savings, or specific technological configurations, while fewer works adopt an integrated techno-economic perspective in which the vertical farm is treated as a coupled agro-energy system. In such systems, the benefits of daylighting cannot be assessed solely in terms of reduced LED electricity consumption, but must also account for crop yield, daily light integral control, supplemental LED operation, HVAC and dehumidification loads, climatic variability, investment costs, and operational costs. Second, despite the potential of daylighting to directly exploit solar radiation and reduce artificial lighting requirements, there is still a lack of critical comparison with more established solar-based alternatives, particularly photovoltaic systems, which can indirectly support vertical farming by supplying electricity to LEDs and auxiliary equipment without modifying the internal light-distribution architecture. Accordingly, this study investigates whether daylighting strategies can offer greater agro-energy and techno-economic advantages than photovoltaic-based solutions for vertical farming. The analysis focuses on identifying the climatic, technical, and economic conditions under which daylighting may become a more convenient option, while ensuring that crop productivity and target light requirements are maintained.

To address these gaps, this study proposes and evaluates an optical-fiber-based daylighting solution for delivering natural light into a small-scale vertical farm. Five optical-fiber configurations are first analyzed through ray-tracing simulations to assess their optical performance and technical feasibility. The resulting daylight availability is then integrated into a previously verified modelling framework for evaluating the agronomic and energy performance of vertical farming systems [16].

Within this framework, the study investigates both fiber-only operation and hybrid fiber–LED strategies, where supplemental artificial lighting is used to meet the target crop light requirements when daylight is insufficient. Two integration approaches are also considered. The first aims to maximize the amount of collectable daylight, while the second is based on the practical constraint of delivering one optical fiber bundle to each growing zone.

The different configurations are compared in terms of specific energy consumption and energy-related costs. Finally, the results are benchmarked against two alternative solar-based strategies: a previously investigated light-pipe system [22] and a photovoltaic-based solution. This comparison allows daylighting and photovoltaic approaches to be assessed within the same techno-economic framework, highlighting the conditions under which optical-fiber daylighting may represent a convenient strategy for reducing the energy burden of vertical farming. Accordingly, the main scientific contributions of this work are summarized as follows.

- Fiber-optic daylighting for vertical farming: development and ray-tracing assessment of five optical-fiber typologies and three integration strategies for delivering natural light into a small-scale vertical farm.
- Agro-energy integration: coupling of ray-tracing results with a validated vertical farming model to evaluate crop light availability, LED supplementation, agronomic performance, and specific energy consumption.
- Techno-economic benchmarking: comparison of fiber-only and hybrid fiber–LED strategies with light-pipe and photovoltaic alternatives, to identify economic trade-offs.

## 2. Methodology

This section presents the assumptions and methods used to assess the techno-economic viability of integrating an optical fiber daylighting system to reduce energy consumption in the vertical farm.

## 2.1. Vertical Farm AGRI-Energy Model

The assessment is carried out using the AGRI-Energy model developed in [16], which is adopted here as the reference framework for simulating the coupled agronomic and energy performance of the small-scale vertical farm. The model links indoor environmental setpoints to both crop growth and subsystem energy demand within the growing chamber. Its inputs include outdoor weather conditions, namely air temperature and solar irradiation, indoor setpoints such as air temperature, relative humidity, $CO_2$ concentration, and photosynthetic photon flux density (PPFD), as well as crop-specific parameters. The model outputs include the power and energy demand of the HVAC system, the water consumption of the vertical farm, and the resulting harvested biomass. By using this previously developed framework, the present study can evaluate how the integration of an optical fiber daylighting system affects environmental control requirements, energy use, and crop productivity.

The chamber energy balance was modified, as shown in Eq. 1, to account for the integration of the optical fiber daylighting system. This was done by introducing an additional thermal contribution, $Q_{OF}$, representing the solar radiative gain delivered to the growing chamber through the optical fibers. All other terms follow the validated AGRI-Energy formulation reported in [16]. In the energy balance, $Q_{LED}$ denotes the radiant power emitted by the LED system, while $Q_{plant}$ represents the fraction of radiation intercepted by the crop and converted through photosynthesis. The sensible loads managed by the heating/cooling system and by the air handling unit are represented by $Q_{h/c}$ and $Q_{AHU}$, respectively. Latent loads related to moisture exchange are described by $Q_{eva}$, associated with crop evapotranspiration, and $Q_{hum}$, associated with humidification. Heat transfer through the chamber envelope is accounted for by $Q_{env}$. The estimation of $Q_{OF}$ is detailed in the next section, whereas the calculation of the remaining energy-balance terms is described in [16].

$$d_{air} c_{p_{air}} V_{room} \frac{\partial T}{\partial t} = Q_{env} + Q_{LED} + Q_{OF} - Q_{plant} - Q_{eva} + Q_{h/c} - Q_{AHU} - Q_{hum} \quad (1)$$

Lettuce was used as the reference crop, given its common application in vertical farming systems and its suitability for controlled-environment production. Crop biomass was estimated at each simulation timestep using the growth sub-model adopted in [16]. This sub-model calculates dry and fresh matter accumulation as a function of air temperature, $CO_2$ concentration, and photosynthetic photon flux density (PPFD), accounting for their effects on leaf area index (LAI), specific leaf area (SLA), and light use efficiency (LUE), as described in Eqs. 2–4. Baseline LUE values were derived from the experimental dataset reported by Carotti et al. [27] and adjusted to account for different $CO_2$ concentrations using the data provided in [28]. The LUE and SLA values obtained from Carotti et al. [27]

and used in this study are summarized in Figure 1~~Figure 1~~. Crop evapotranspiration was calculated consistently with the approach described in [16]. Since the crop-growth and evapotranspiration routines were not modified in the present work, only their main assumptions are reported here, while the full formulation is provided in [16].

$$\frac{\partial DM}{\partial t} = PPFD \cdot (1 - e^{-k \cdot LAI}) \cdot LUE_{dm} \cdot A_{crop} \quad (2)$$

$$\frac{\partial LAI}{\partial t} = \frac{\partial DM}{\partial t} \cdot SLA \quad (3)$$

$$\frac{\partial FM}{\partial t} = PPFD \cdot (1 - e^{-k \cdot LAI}) \cdot LUE_{fm} \cdot A_{crop} \quad (4)$$

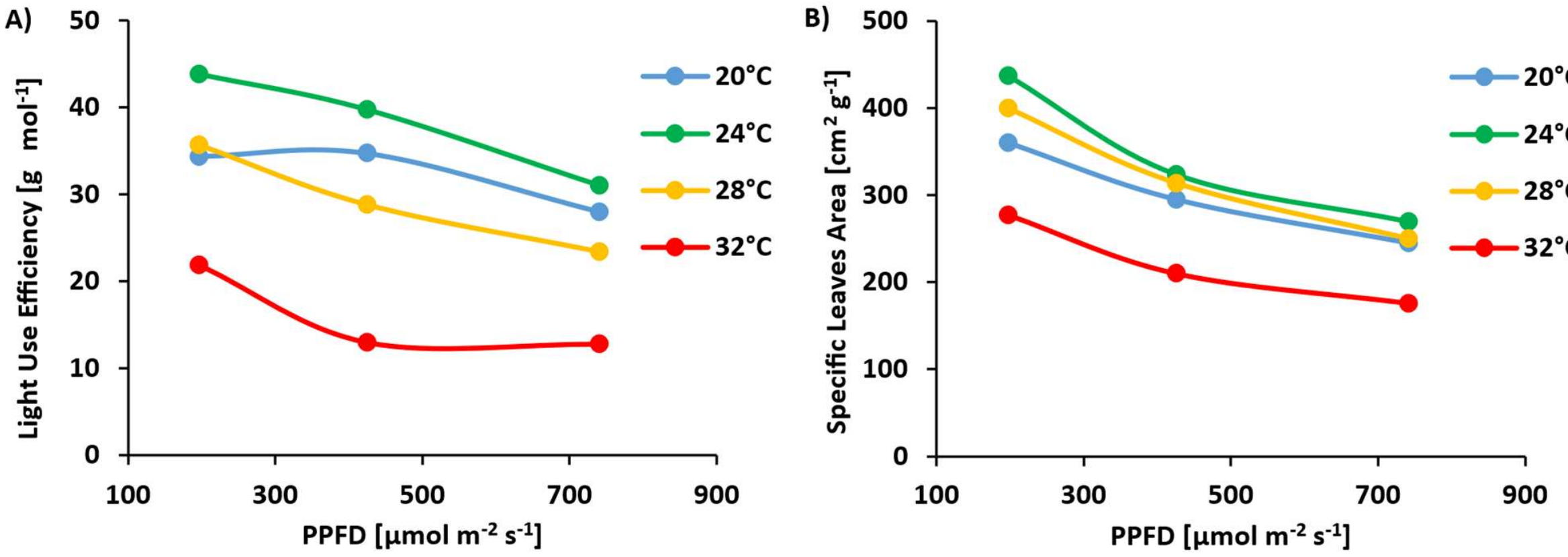

Figure 1: Experimental Light use efficiency (A) and Specific leaves area (B) of lettuce under different growing conditions.

## 2.2. Proposed Vertical Farm Configuration

The case study refers to a small-scale vertical farm with external dimensions of 7 × 7 × 3 m, representative of compact systems suitable for urban applications. To limit heat transfer with the outdoor environment, the farm envelope is assumed to consist of polyurethane sandwich panels, with a U-value in the range of 0.17–0.18 W $m^{-2}$ $K^{-1}$. The internal layout includes five growing racks, each arranged over three cultivation tiers, resulting in a total growing area of 90 $m^2$ [16], calculated as the sum of the cultivation surfaces across all tiers. This layout, together with the main design assumptions, is kept consistent with the previous light-pipe study [22] in order to enable a direct comparison between the two daylighting technologies.

Lettuce is considered as the reference crop and is grown at a planting density of 25 plants $m^{-2}$, with harvesting assumed at an average fresh mass of 250 g per plant, consistently with the experimental data reported in [27]. The indoor environmental setpoints are selected according to the sensitivity analysis presented in [11], adopting the configuration associated with the lowest levelized cost of lettuce production. Relative humidity setpoints are instead defined mainly to reduce the risk of condensation and microbial development, given their limited effect on biomass accumulation. The vertical farm is assumed to be located in Dubai, where high solar availability makes the integration of daylighting strategies particularly relevant. Moreover, in arid regions such as the United Arab Emirates, controlled-environment systems with reduced freshwater requirements can contribute to food production under water-scarce conditions. In the analyzed configuration, water-use efficiency is enhanced by recovering condensate from the HVAC cooling coils and recirculating it within the system, so that the net water demand is largely associated with the water retained in the harvested biomass. The complete set of design parameters and operating setpoints is reported in Table 1Table 1.

Table 1: VF design parameters

| Variable | Value | Unit | Reference |
|---|---|---|---|
| **Vertical Farm Area** | 49 | $m^2$ | [16] |
| **Cultivation Area per Tier** | 30 | $m^2$ | [16] |
| **Overall Cultivation Area** | 90 | $m^2$ | [16] |
| **Crop Density** | 25 | plants $m^{-2}$ | [27] |
| **Envelope Thermal Transmittance** | 0.17 – 0.18 | W $m^{-2}$ $K^{-1}$ | [16] |
| **Indoor Air Temperature Setpoint** | 24 | °C | [11] |
| **$CO_2$ Level Setpoint** | 1,200 | ppm | [11] |
| **PPFD Setpoint** | 250 | µmol $m^{-2}$ $s^{-1}$ | [11] |
| **Relative Humidity Setpoint** | 75% (light) / 85% (dark) | - | [27] |

This baseline configuration is used to quantify the impact of the optical fiber system on the chamber energy balance and overall VF performance.

### 2.3. Proposed Daylighting Configuration

The main advantage of an optical fiber-based daylighting system is its ability to deliver natural light directly to selected growing zones, including individual cultivation layers in multilayer vertical farming structures. This feature is particularly relevant in stacked systems, where direct solar access is limited and the distribution of natural light among tiers is intrinsically non-uniform. Previous studies on daylighting and optical fiber systems have shown that solar concentrators can improve system performance by increasing the amount of daylight collected and transmitted indoors [29] [24].

In the present study, the optical fiber daylighting system is coupled with a roof-mounted linear concentrating unit, schematically shown in Figure 2Figure 2. The figure reports two views of the same device. The transverse view illustrates the focusing geometry of the Fresnel lens and the position of the optical receiver, while the longitudinal view shows the linear development of the receiver and the corresponding optical fiber bundle. The concentrating unit consists of a long linear Fresnel lens supported by a rigid frame and aligned with a linear optical receiver positioned along the focal region of the lens. Incident solar radiation is concentrated onto the receiver, coupled into a bundle of optical fibers, and then delivered to the cultivation area.

Unlike the experimental configuration proposed by Vu et al. [24], where the delivered PPFD was constrained by the limited prototype scale, the proposed design assumes that the available roof surface of the vertical farm is used to accommodate a larger collecting area. The system is therefore conceived as a modular roof-integrated solution, in which one or more linear Fresnel units can be installed according to the available surface and the required light input. The concentrating unit is based on the Linear Fresnel lens concept investigated in [29], where a flat elongated Fresnel lens focuses incident sunlight onto a linear receiver.

At this preliminary design stage, a linear Fresnel lens was selected instead of a parabolic concentrator mainly to simplify the optical layout and the solar-tracking mechanism. The proposed system performs tracking with respect to the solar altitude angle by rotating the whole lens–receiver assembly around a longitudinal shaft. In this way, the relative position between the Fresnel lens and the focal receiver remains fixed during operation, while the inclination of the unit is adjusted to follow the apparent daily motion of the sun. This solution avoids the need for the independent movement or repeated optical realignment of multiple components.

The rotation is driven by two lateral motors placed at the opposite ends of the shaft. This configuration has a dual function. First, it provides the actuation required for solar tracking along the solar-altitude axis. Second, it reduces the torsional load that would occur if the long concentrating unit were driven from only one side. By applying torque at both ends, the mechanical stress on the shaft and supporting frame is limited, improving the structural behavior of the system and reducing the risk of misalignment between the lens and the receiver. This aspect is particularly important because the unit has an elongated geometry, and even small torsional deformations could affect the optical coupling efficiency.

For the same reason of mechanical and optical simplification, the collimating lens and trough compound parabolic concentrator included in the configuration proposed by Ullah et al. [29] were not considered in the present design. Although this simplified configuration may result in a lower concentration efficiency compared with more complex multi-stage optical systems, it reduces the number of components requiring alignment and makes the system more suitable for preliminary integration on the roof of a vertical farm.

Finally, the Fresnel lens is arranged with its grooved surface facing downward, as shown in Figure 2~~*Figure 2*~~. This orientation leaves the upper exposed surface smooth, facilitating cleaning and reducing dust accumulation within the lens grooves. This design choice is particularly relevant for installations in dusty climates, such as Dubai, where soiling can significantly reduce the optical performance of concentrating systems. Overall, the proposed configuration should therefore be interpreted as a preliminary design solution aimed at balancing daylight collection, solar-tracking capability, mechanical simplicity, maintainability, and suitability for modular integration on vertical farms. The following sections describe the design procedure adopted to size the overall system and define its main optical and mechanical components.

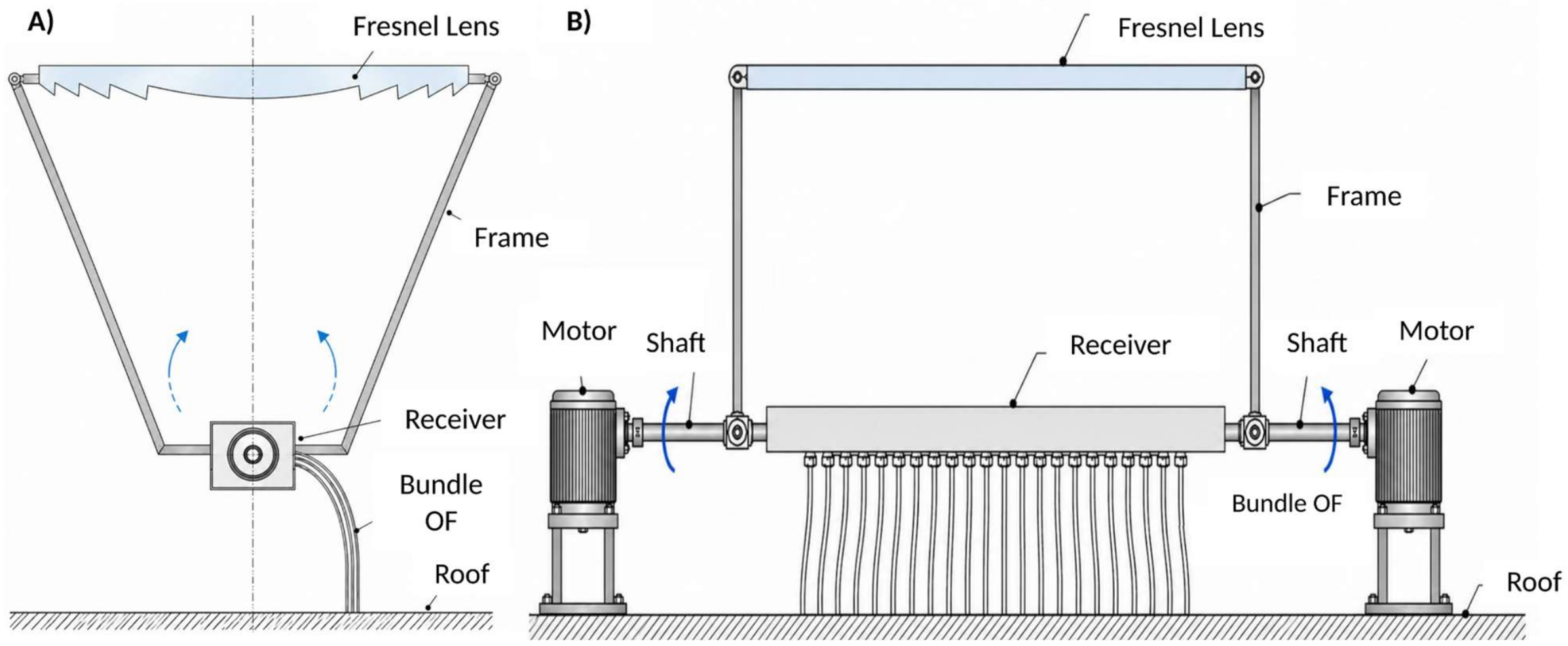

Figure 2: Schematic representation of the proposed roof-mounted optical fiber daylighting concentrator: A) transverse view of the Fresnel lens, focal receiver, and optical fiber bundle; B) longitudinal view of the linear receiver, fiber bundle distribution, shaft, and motorized tracking system. The Fresnel lens grooves face downward to reduce soiling and facilitate cleaning.

### 2.3.1. Fresnel Lens Parameters Selection

The first step in the design of the proposed system was the selection of the Fresnel lens, with the objective of maximizing the amount of daylight concentrated onto the optical fiber bundles while maintaining a mechanically feasible configuration for solar tracking. The main design parameters of a Fresnel lens are the aperture width (W), the focal length (f), and the groove pitch. In the present study, the lens aperture was set to 500 mm. This value is smaller than those adopted in comparable daylight-concentrating systems, but it was selected to limit the size and weight of the movable assembly and thus reduce the mechanical effort required for tracking.

The groove pitch was set to 5 mm, which is larger than the value adopted by Yang et al. [30]. This choice was made to increase the mechanical robustness of the lens and make it more suitable for prolonged outdoor operation. Although a smaller groove pitch could improve optical accuracy, it would also make the lens more delicate and potentially more sensitive to manufacturing defects, dust accumulation, and degradation over time. In addition, the central portion of the lens was designed as a smooth convex region without grooves, with a width of 50 mm. This geometry was adopted to provide a continuous central focusing surface while preserving the grooved Fresnel structure in the lateral portions of the lens.

The focal length was then selected according to the optical acceptance of the fiber bundle. The amount of light that can be effectively coupled into the fibers depends on the numerical aperture of the optical fibers (NA), which is determined by the refractive indices of the fiber core and cladding, as expressed in Eq. (5). This acceptance angle was compared with the maximum semi-angle of the Fresnel lens, defined in Eq. (6), which represents the angular spread of the rays focused by the lens toward the receiver. To ensure efficient optical coupling, the focal length was chosen so that the Fresnel lens semi-angle remained lower than the acceptance angle of the optical fibers.

$$NA = \sin^{-1}(\sqrt{n_{core}^2 - n_{clad}^2}) \quad (5)$$

$$\nu_{lens} = \tan^{-1}(\frac{W/2}{f}) \quad (6)$$

However, the focal length could not be increased arbitrarily, since a longer focal length would require a larger lens–receiver distance and, consequently, a heavier and more complex tracking structure. Therefore, the selected focal length was the minimum value able to satisfy the optical coupling condition while preserving a compact and mechanically feasible configuration.

#### 2.3.2. Optical Fiber Thermal Balance

A thermal balance model was developed to support the selection of the receiver position with respect to the focal point. The model evaluates the effect of the concentrated solar flux on the temperature reached at the optical fiber surface under high-intensity daylighting conditions. This analysis provides a first-order thermal criterion for identifying receiver positions that ensure suitable operating temperatures while preserving the optical concentration performance of the system. The thermal model was intentionally simplified in accordance with the scope of the study and was used to identify a representative receiver position and geometry for the subsequent techno-economic analysis. More detailed modelling would be required at the final engineering design stage.

An optical fiber with an external diameter of 1 mm was considered. The fibers were arranged in a hexagonal packing configuration, as schematically shown in Figure 3~~Figure 3~~. Each fiber was described by its main material regions, namely the core, the cladding, and the surrounding filler material. This arrangement was selected to increase the effective surface available for daylight collection and transport within a compact receiver geometry.

To describe the thermal interaction between the fiber and the surrounding material, the filler region associated with the hexagonal packing was represented by an equivalent cylindrical layer surrounding the cladding, as shown in Figure 3~~Figure 3~~. The equivalent layer was defined by preserving the same filler cross-sectional area. This geometrical representation allows the heat transfer through the filler to be expressed by means of radial conductive thermal resistances.

The thermal balance was applied to a representative cylindrical control volume with a thickness of 5 cm, as illustrated in Figure 3~~Figure 3~~. For each material region, namely core, cladding, and filler, a steady-state energy balance was formulated. The equilibrium temperature of each region was evaluated by considering the incoming concentrated solar power ($Q_{in}$), the transmitted optical power ($Q_{tr}$), the absorbed power ($Q_{abs}$), and the heat exchanged through conduction ($Q_{cond}$), convection ($Q_{conv}$), and thermal radiation ($Q_{rad}$).

The concentrated heat flux distribution on the receiver plane was obtained through ray-tracing simulations performed with the Tonatiuh software. Further details on the Tonatiuh ray-tracing setup are reported in Appendix 7.8. Following the procedure reported in [31], the Fresnel lens and the receiver were reproduced in the simulation environment to estimate the maximum heat flux at the center of the receiver for different axial displacements with respect to the focal length. The resulting flux values were then used as input data for the thermal balance model. The analysis was carried out for an incident solar irradiance of 800 W $m^{-2}$, representative of high-intensity daylighting operating conditions at the selected location and therefore suitable for preliminary receiver sizing, since only the direct component is concentrated by the Fresnel lens. A convective heat-transfer coefficient of 25 W $m^{-2}$ $K^{-1}$ was adopted as a representative forced-convection value.

The central optical fiber was selected as the reference fiber for the thermal analysis, since it is subjected to the highest concentrated flux. The six surrounding fibers were assumed to operate under the same thermal condition as the central one. Accordingly, the fiber bundle was treated as a locally symmetric domain, and the heat transfer problem was reduced to a radial balance applied to the representative fiber and its equivalent surrounding filler layer.

The filler region was modelled as an equivalent cylindrical layer to compute the conductive thermal resistance between the cladding and the surrounding medium. Since each filler portion in the hexagonal packing is shared by three adjacent fibers, the heat exchanged through the filler was multiplied by a factor of three. This correction accounts for the thermal interaction between the central fiber, the filler, and the neighboring fibers within the packed bundle.

Under these assumptions, the steady-state energy balance for the core is expressed as shown in Eq. (7), where $Q_{in,core}$ is the concentrated power incident on the core, $Q_{tr,core}$ is the optical power transmitted through the core, $Q_{abs,core}$ is the power absorbed by the core material, $Q_{conv,core}$ and $Q_{rad}$ are the convective and radiative heat losses, respectively, and $Q_{cond,core\text{-}clad}$ represents the conductive heat exchange between the core and the cladding. Eqs. (8-9) represent the corresponding balance for cladding and filler layer, respectively.

$$Q_{in,core} = Q_{tr,core} + Q_{conv,core} + Q_{rad,core} + Q_{cond,core-clad} \quad (7)$$

$$Q_{in,clad} = Q_{tr,clad} + Q_{conv,clad} + Q_{rad,clad} + Q_{cond,core-cla} + Q_{clad-filler} \quad (8)$$

$$Q_{in,filler} = Q_{tr,filler} + Q_{conv,filler} + Q_{rad,filler} + Q_{cond,clad-fille} + Q_{filler-ext} \quad (9)$$

For all the balances, conductive terms are defined according to the temperature gradient between adjacent regions and are considered positive when heat is transferred from the control volume under analysis toward the neighboring region. The material properties and the detailed expressions used to compute the optical, conductive, convective, and radiative terms are reported in Appendix 7.3.

The proposed thermal balance allows the equilibrium temperature of the optical fiber surface to be estimated as a function of the receiver displacement from the focal point. This temperature-based criterion was then used to identify the receiver positions that combine effective daylight concentration with suitable thermal operating conditions for the optical fiber bundle.

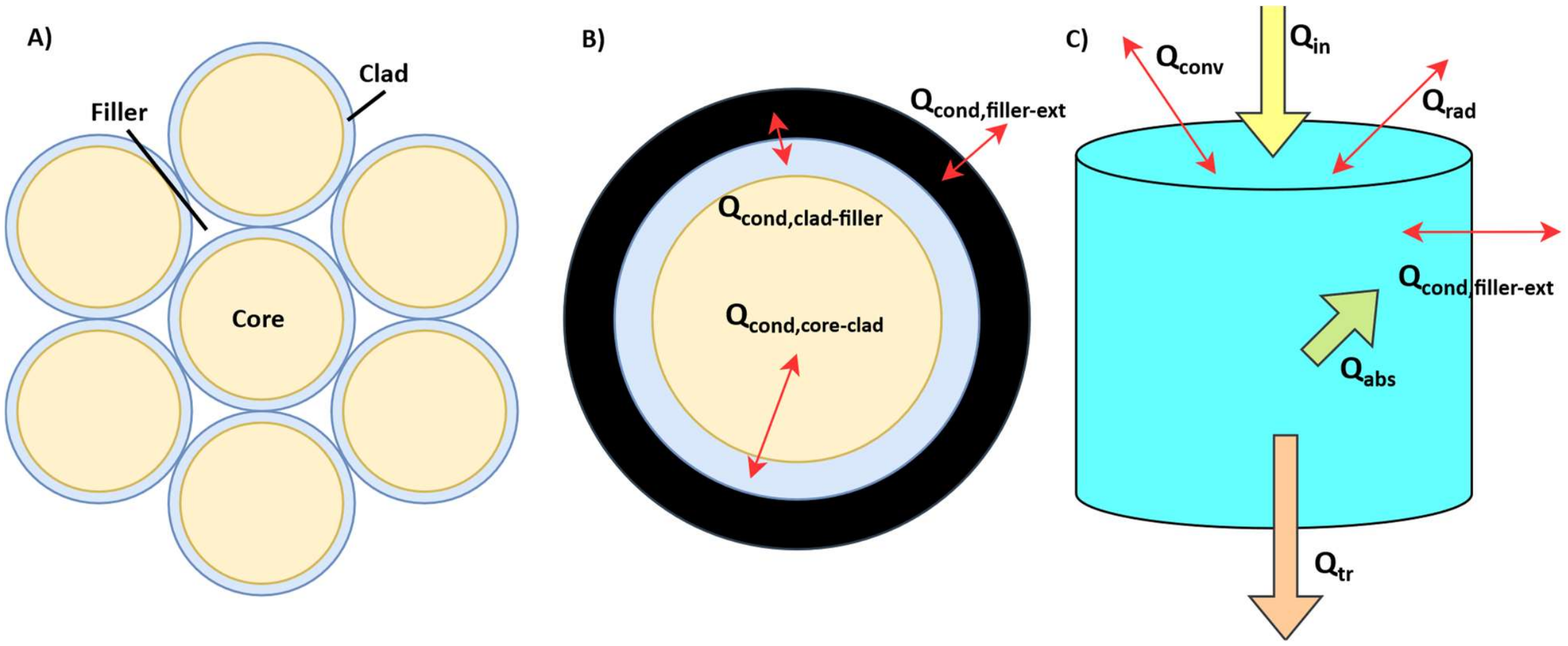

Figure 3: Schematic representation of the optical fiber thermal balance model: hexagonal packing of the fibers and filler region (A), equivalent multilayer cylindrical geometry used to evaluate radial conductive resistances (B), and energy balance terms considered for the representative control volume, including incoming, transmitted, absorbed, convective, radiative, and conductive heat fluxes (C).

### 2.3.3. Collector Orientation and Tracking strategy

Since the selected concentration system allows only monoaxial solar tracking, the collector orientation was optimized to maximize the annual amount of concentrated solar energy. For this purpose, the collected direct solar energy was estimated using Eq. (10), where $I_{dir}$ is the direct solar irradiance obtained from the PVGIS database for the selected location [32], and cos(θ) accounts for the incidence angle of sunlight on the tilted collector surface.

$$I_{Fresnel} = I_{dir} \cdot \cos\theta \tag{10}$$

The incidence factor cos(θ) was evaluated as a function of the location-specific solar position, described by the solar azimuth ($\gamma_{sol}$) and solar elevation angles ($\alpha_{sol}$), and of the collector orientation, described by its azimuth ($\gamma_{Fresnel}$) and tilt angle ($\alpha_{Fresnel}$). The solar position angles were calculated according to the procedure reported in Appendix 7.2, while the relationship between the solar position and the collector orientation is expressed in Eq. (11).

$$\cos\theta = \sin\alpha_{sol} \cdot \cos\alpha_{Fresnel} + \cos\alpha_{sol} \cdot \sin\alpha_{Fresnel} \cdot \cos(\gamma_{sol} - \gamma_{Fresnel}) \tag{11}$$

Within the monoaxial tracking configuration, the collector tilt angle was adjusted as a function of the instantaneous solar elevation and solar azimuth, as defined in Eq. (12). Therefore, the tracking strategy allows the collector's tilt to vary during operation, while the collector azimuth remains fixed. An optimization procedure was then carried out to identify the fixed azimuth orientation that maximizes the annual collected direct solar energy over the entire year.

$$\alpha_{Fresnel} = \tan^{-1}(\tan(\frac{\pi}{2} - \alpha_{sol}) \cdot \cos(\gamma_{sol} - \gamma_{Fresnel})) \tag{12}$$

To maximize the solar energy collected on the farm roof, multiple linear Fresnel concentrators were considered on the available flat roof surface. The collector azimuth orientation was defined according to the procedure described in the previous section, while the center-to-center spacing between adjacent concentrators was selected to maximize the annual collected direct solar energy.

The spacing between consecutive concentrators is a key design parameter, as it affects both the optical performance of each unit and the number of concentrators that can be installed on the available roof area. A sensitivity analysis was therefore carried out to identify the best trade-off between concentrator density and mutual shading losses. Smaller spacing values allow a larger number of concentrators to be placed on the roof, increasing the total aperture area. However, they can also lead to partial shading between adjacent units, reducing the effective illuminated aperture of each lens. Conversely, larger spacing values reduce shading losses, but decrease the number of installable concentrators.

The focal length of the Fresnel lens was also included in the geometrical assessment of the concentrator layout. Since the receiver position is linked to the focal distance, larger focal lengths increase the overall height of the concentrating module and, consequently, the minimum spacing required to limit mutual shading between consecutive collectors.

The sensitivity analysis was performed on an hourly basis over the entire year. For each time step, the instantaneous collector tilt, determined by the monoaxial tracking strategy, was used to evaluate the mutual shading condition between consecutive concentrators. The corresponding reduction in effective lens aperture was then calculated and included in the estimation of the collected direct solar energy. The detailed geometrical formulation used to determine the apparent collector height, the minimum shading-free spacing, the ideal center-to-center distance, and the shaded lens portion is reported in Appendix 7.4.

For each assumed center-to-center spacing, the number of concentrators that could be installed on the available roof surface was determined. The annual collected energy was then calculated by combining the number of installed concentrators with the hourly effective aperture of each lens. The optimal spacing was identified as the configuration that maximizes the total annual collected direct solar energy, while accounting for mutual shading, concentrator density, and the geometrical constraints imposed by the focal length.

### 2.4. Optical transmission chain

To estimate the PPFD delivered by the optical fiber system to the crop growing zone, the losses occurring along the entire optical transmission chain were considered. As expressed in Eq. (13), the direct irradiance $I_{dir}$ was first projected onto the tilted concentrator surface through the incidence factor. The term $\tau_{Fresnel}$ accounts for the transmittance of the Fresnel lens, assumed equal to 0.92 [24], while $\eta_{conc}$ represents the concentration efficiency obtained from the Tonatiuh ray-tracing simulations. The coefficient $\tau_{UV\text{-}IR}$, assumed equal to 0.455 [24], describes the transmittance of the UV-IR filter, which is introduced to remove the spectral components outside the useful range (PAR) and to reduce overheating caused by undesired wavelengths. The packing efficiency $\eta_{pack}$, equal to 0.87, accounts for the fraction of the receiver area effectively occupied by the optical fibers arranged in the hexagonal bundle. The term $\eta_{inlet}$ represents the optical losses occurring at the fiber inlet during the coupling between the focused radiation and the optical fiber bundle. The diffuser transmittance $\tau_{diff}$, assumed equal to 0.92 [22], accounts for the light transmitted through the terminal diffuser used to distribute the delivered radiation over the cultivation area. The bundle efficiency $\eta_{bundle}$ includes additional generic losses associated with the optical fiber bundle, while $\eta_{tr}$ represents the transmission efficiency of the fibers along their length. Finally, $f_{dl}$ is the conversion factor used to express the delivered daylight from W m$^{-2}$ to µmol m$^{-2}$ s$^{-1}$, accounting for the fact that the non-PAR fraction has already been removed by the UV–IR filter, and was assumed equal to 4.57 µmol J$^{-1}$ [33]. Further details on the assumptions and coefficients adopted to describe the optical transmission chain are reported in Appendix 7.4.

$$PPFD_{OF} = I_{dir} \cdot \cos\theta \cdot \eta_{shade} \cdot \tau_{Fresnel} \cdot \eta_{conc} \cdot \tau_{UV-I} \cdot \eta_{pack} \cdot \eta_{inlet} \cdot \tau_{diff} \cdot \eta_{bundle} \cdot \eta_{tr} \cdot \frac{A_{Fresnel}}{A_{cult}} \cdot f_{dl} \quad (13)$$

### 2.5. PV strategy

To compare the OF-based daylighting strategy with a conventional use of the available VF roof surface, a reference PV system was also evaluated from both economic and thermodynamic perspectives. The PV electrical output was estimated using Eq. (14), which accounts for the direct and diffuse components of the incident solar radiation.

$$P_{PV} = (I_{dir} \cdot \cos\theta \cdot (1 - \rho_{dir}) + I_{diff} \cdot (\frac{1+\cos\ \ _{PV}}{2}) \cdot (1 - \rho_{diff})) \cdot \eta_T \cdot \eta_{inv} \cdot \eta_{BOS} \quad (14)$$

In the equation, $I_{dir}$ and $I_{diff}$ represent the direct and diffuse irradiance, respectively. The term cos(θ) accounts for the incidence angle of direct radiation on the tilted PV surface, while the coefficients $\rho_{dir}$ and $\rho_{diff}$ describe the optical response of the PV surface to direct and diffuse radiation, respectively. Finally, $\eta_T$, $\eta_{inv}$, and $\eta_{BOS}$ represent the temperature correction, inverter efficiency, and balance-of-system efficiency. The detailed formulations of these efficiency terms are reported in Appendix 7.5. In the PV model, the technical characteristics of an existing photovoltaic panel were adopted from [34], while the approaches proposed in [35], and [36] were used to evaluate the module efficiency and its dependence on operating conditions. The additional efficiency terms included in the PV performance chain were assessed following the methods reported in [37], and [38].

In this study, an additional frame structure is assumed to be located on the rooftop of the vertical farm in order to fix an optimal tilt angle for the PV panels avoiding the complexity of a solar-tracking system, and the necessity of distancing the modules to prevent mutual shading.

### 2.6. Economic Assumptions

An economic assessment was carried out to complement the energy analysis of the proposed daylighting strategy. The potential reduction in artificial lighting demand was evaluated together with the capital and operating costs required for the implementation of the optical-fiber system, with the aim of identifying the conditions under which the proposed solution could represent a realistic and reasonable investment for small-scale vertical farming applications.

For each investigated configuration, capital and operating costs were estimated according to the economic assumptions summarized in Table A3 in the Appendix, including the capital costs of the Fresnel lens [24], optical fibers [39][40][41][42] and PV system components [43][44][45]. Tracking electricity consumption, diffuser costs, and POF degradation were intentionally excluded because of the uncertainty associated with their estimation at this preliminary design stage. The economic results are strongly dominated by the purchase cost of the optical fibers. Therefore, including these additional contributions would further penalize the daylighting configurations without altering the main conclusions of the analysis. The investigated configurations were compared using two simplified economic indicators: the cost per unit of PAR delivered at crop level and the simple payback time. The first indicator relates the total cost of the system to the useful radiation effectively supplied to the plants, allowing the cost-effectiveness of the different daylighting configurations to be compared on the basis of the light actually delivered to the crop.

The simple payback time was calculated by considering the annual economic benefit associated with the proposed strategy. This benefit includes the reduction in electricity costs due to the lower artificial lighting demand, the avoided carbon emissions valued through an assumed carbon tax, and the variation in vertical farm productivity compared with the reference configuration [7]. Therefore, the payback time accounts not only for direct energy-cost savings, but also for the economic effect of emission reduction and possible changes in crop production.

Given the preliminary techno-economic scope of this study, the analysis focuses on the main cost and performance drivers affecting the convenience of the proposed strategy. The comparison is therefore based on non-discounted indicators, without including discount-rate assumptions, net present value, or discounted payback time.

## 3. Investigated Configurations

To investigate the potential integration of an optical-fiber-based daylighting strategy for reducing the lighting energy demand of the vertical farm, several configurations were defined and analyzed. These configurations include different optical fiber typologies and integration strategies, allowing the influence of both component selection and system layout on the overall performance to be assessed.

### 3.1. Optical Fibers

In this study, five optical fiber typologies were investigated. Since the proposed system is designed to transmit lighting power rather than information signals, optical fibers with diameters smaller than 1 mm were excluded from the analysis. The selected fibers include both polymer optical fibers (POFs) and silica optical fibers (SOFs). POFs are generally characterized by higher flexibility, easier handling, and lower cost, but they are more sensitive to thermal constraints due to their lower maximum operating temperature. Conversely, SOFs typically provide higher thermal stability and lower optical attenuation, although they are associated with higher costs and lower mechanical flexibility.

Each optical fiber is characterized by its diameter, numerical aperture, attenuation coefficient, maximum operating temperature, and cost. The numerical aperture depends on the refractive indices of the core and cladding and determines the acceptance angle of the fiber, while the attenuation coefficient describes the optical losses occurring during light propagation along the fiber length. The main properties of the investigated optical fibers are summarized in Table 2~~Table 2~~.

Table 2: Main properties and bundle configurations of the optical fibers analyzed in this study. The attenuation coefficient adopted for each fiber was selected as a representative value over the PAR spectral range.

| Parameter | POF1 | POF2 | POF3 | SOF1 | SOF2 | Unit |
|---|---|---|---|---|---|---|
| **Diameter** | 1.0 | 1.0 | 3.0 | 1.0 | 2.0 | mm |
| **Core Index** | 1.494 | 1.492 | 1.492 | 1.457 | 1.457 | [-] |
| **Clad Index** | 1.460 | 1.402 | 1.402 | 1.411 | 1.411 | [-] |
| **NA** | 0.322 | 0.535 | 0.535 | 0.372 | 0.372 | [-] |
| **Max T** | 80.0 | 80.0 | 80.0 | 125.0 | 125.0 | °C |
| **Attenuation** | 0.250 | 0.250 | 0.250 | 0.042 | 0.042 | dB $m^{-1}$ |
| **Bundle Width** | 20.0 | 24.0 | 24.0 | 10.0 | 10.0 | mm |
| **OF per Bundle** | 461 | 665 | 73 | 115 | 28 | [-] |

For each case, the optical fibers were arranged in a square-shaped bundle using a hexagonal packing distribution to maximize the effective surface available for light transmission. The width of the square bundle was defined to ensure high concentration efficiency, allowing most of the radiation focused by the Fresnel lens to be intercepted by the receiver. The receiver size, the distance between the Fresnel lens and the receiver, and the spacing between adjacent concentrators were evaluated for each optical fiber according to the procedures described in Sections 2.3.2 and 2.3.3. The resulting design parameters are presented in the Results section.

### 3.2. Integration Strategies

For each optical fiber typology, three integration strategies were investigated. First, a natural-light-only configuration was evaluated, in which the artificial lighting system is completely replaced by the optical-fiber daylighting system. This configuration is referred to as NL. Second, a hybrid lighting configuration was considered, in which the optical-fiber system is coupled with supplemental LED lighting. This configuration is referred to as HS. In the hybrid strategy, the LED system is operated to ensure a target PPFD of 250 μmol $m^{-2}$ $s^{-1}$ over the required 16 h photoperiod when the daylight contribution is insufficient. The nominal photosynthetic photon efficiency of the assumed LED system was 2.5 μmol $J^{-1}$.

The operation of the LED system was also constrained by the characteristics of the PWM-modulated driver. In particular, the LEDs were assumed to operate only above a minimum power load ratio of 30%. Therefore, when the required supplemental lighting power falls below this threshold, the LED system cannot be further dimmed and the corresponding operating constraint is applied in the model.

Two different bundle integration approaches were then defined. The first approach, named MAX, aims to maximize the amount of daylight collected by exploiting all the optical fiber bundles that can be installed on the available roof surface. In this case, the collected daylight is distributed among all the growing zones. This approach was adopted for the NL configuration, where the maximization of collected natural light is essential to limit crop light deficits and associated productivity losses. The same approach was also evaluated in combination with the hybrid LED–OF strategy.

The second approach, named MIN, represents a more conservative configuration in terms of the number of optical fibers and installed daylighting components. In this case, each growing zone is served by one optical fiber bundle. Following the experimental setup proposed by Carotti et al. [27], each growing zone was assumed to be a square area of 200 mm side length. Considering a cultivated area of 90 $m^2$, this corresponds to 750 growing zones per cultivation layer.

Overall, the analyzed strategies include the natural-light-only configuration with maximum daylight collection, the hybrid configuration with maximum daylight collection, and the hybrid configuration with one bundle per growing zone.

## 4. Results

### 4.1. Daylighting System

The coupling between the linear Fresnel concentrator and the optical fiber receiver was assessed by considering the maximum temperature reached by the fiber bundle at the receiver inlet. This criterion was adopted to identify receiver positions that provide effective optical coupling while maintaining suitable thermal operating conditions for the selected optical fibers. To determine the concentrated flux entering the receiver, a sensitivity analysis was performed using the Tonatiuh ray-tracing software. The analysis was carried out by varying the axial displacement of the receiver with respect to the nominal focal point of the Fresnel lens, indicated as (Z). For each receiver position, the solar flux profile along the receiver width was obtained and the maximum concentrated flux was used as input for the thermal balance model. As shown in Figure 4Figure 4, increasing the distance from the focal point leads to a broader and less intense flux distribution on the receiver surface. This trend was observed for both positive and negative displacements with respect to the focal position, indicating that moving the receiver away from the focal point reduces the peak thermal load on the optical fiber bundle. Therefore, the receiver position was selected by balancing the need for high concentration efficiency with the requirement of limiting the maximum fiber temperature. For clarity, only the results obtained for the POF1 configuration are reported in this section. The same procedure was applied to the other optical fiber typologies. Slight differences in the resulting flux profiles and optimal receiver positions are expected among the investigated configurations due to the different focal lengths associated with the selected Fresnel lens configurations.

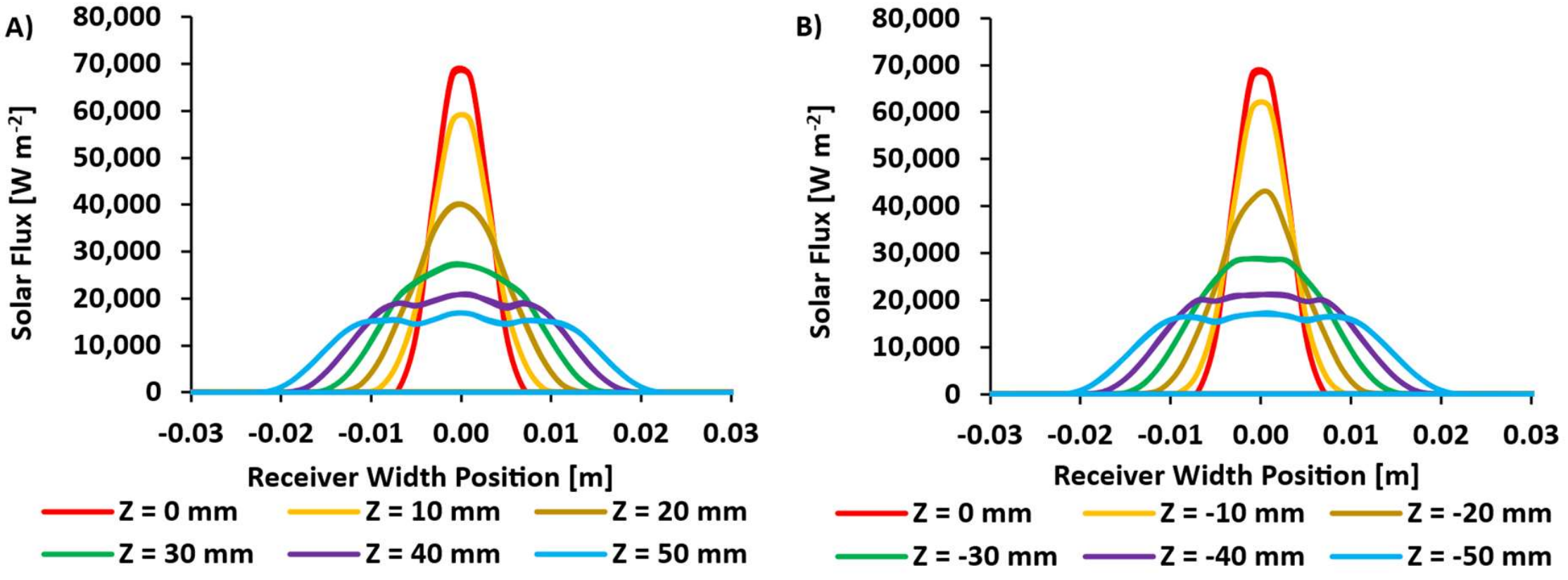

Figure 4: Solar flux profiles along the receiver width at different axial positions, showing peak attenuation and profile broadening as Z increases.

Based on the heat flux distributions obtained from the sensitivity analysis performed with Tonatiuh, the maximum temperature profiles reported in Figure 5~~Figure 4~~ were evaluated for each configuration. The POF3 and SOF2 trends are not shown, since their deviations from POF2 and SOF1, respectively, were found to be negligible.

As expected, the maximum temperature of the optical fiber bundle increases as the receiver approaches the focal position. Among the polymer optical fiber configurations, POF2 reaches the highest temperature at the focal point, mainly due to the higher incident heat flux associated with its shorter focal length, approximately 56% of that of POF1. Although the SOF configuration is exposed to a higher heat flux than POF1, its lower attenuation factor reduces the fraction of radiation absorbed within the first layer of the fiber bundle. As a result, the maximum temperature reached by SOF remains comparable to that of POF1 and well below its operating limit.

The optimal receiver position was therefore selected by ensuring that the maximum temperature remained below the allowable operating threshold of each optical fiber type, indicated by the dashed horizontal lines in Figure 5~~Figure 5~~. For POF1 and POF2, axial displacements of 15.0 mm and 10.0 mm were selected, respectively, to prevent the corresponding thermal limits from being exceeded. Conversely, owing to its lower attenuation factor and higher allowable operating temperature, the SOF configuration can be positioned directly at the focal point.

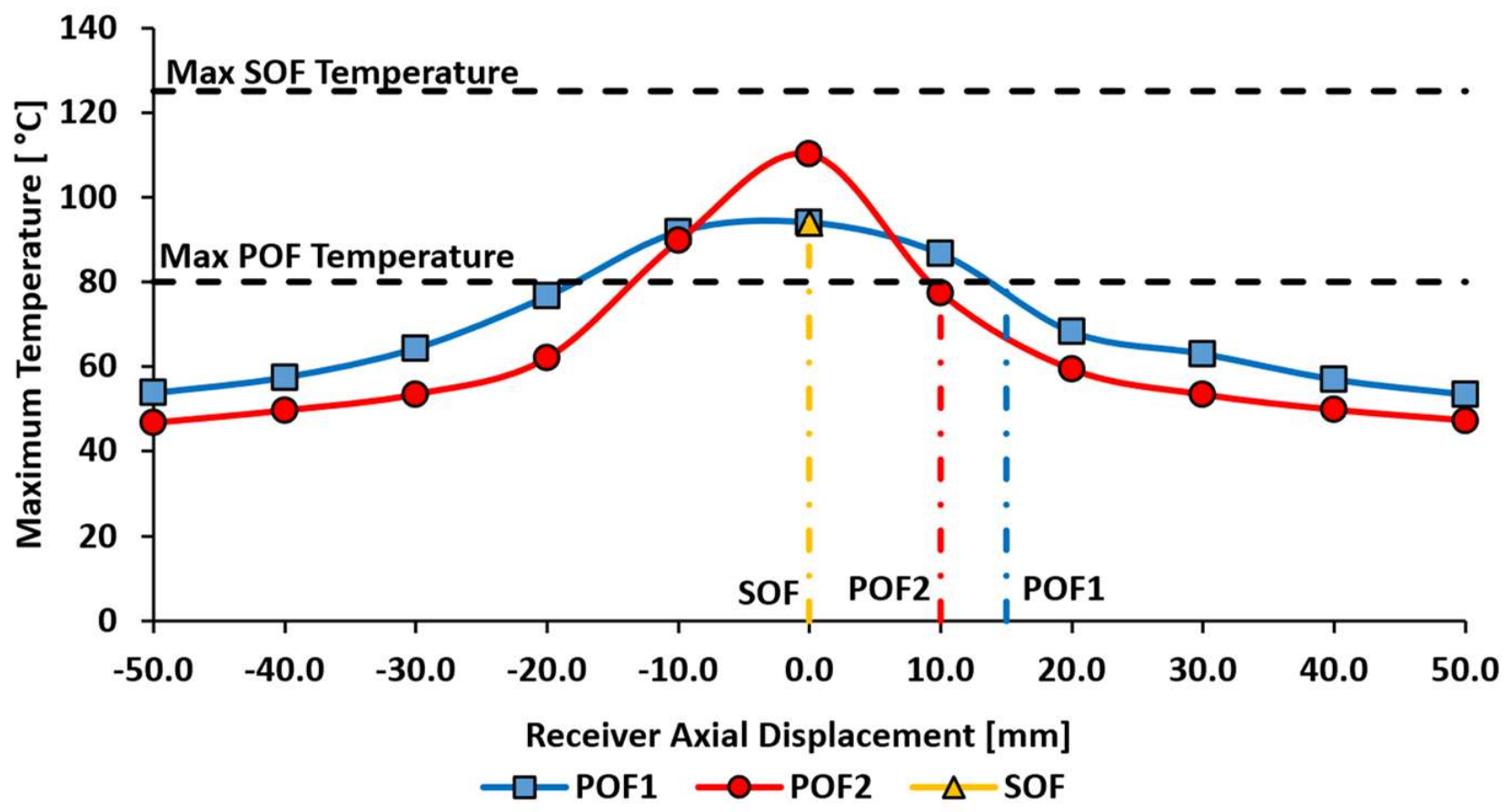

Figure 5: Maximum temperature profiles of the optical fiber bundle as a function of receiver axial displacement for the POF1, POF2, and SOF configurations. The dashed lines indicate the maximum allowable operating temperatures for polymer and silica optical fibers, while the vertical markers identify the selected receiver positions.

The receiver width was finally assessed by considering the amount of daylight energy that could be effectively captured. To achieve high concentration efficiency and limit optical losses, the receiver size was selected according to the daylight flux distribution across the receiver width for each optical fiber configuration, evaluated at the optimal axial displacement identified in the previous section. The flux profiles reported in Figure 6~~Figure 6~~ were obtained using Tonatiuh.

Although the flux distributions are not uniform, the selected receiver widths, 20 mm for POF1, 24 mm for POF2 and POF3, and 10 mm for SOF1 and SOF2, allow most of the incident energy to be collected while avoiding excessive concentration losses. The non-uniform flux distribution across the bundle is expected to affect only the relative contribution of each fiber to the total transmitted light. After collection, each bundle is connected to the growing zone, where the transmitted light is redistributed by a diffuser. Therefore, an acceptable light distribution can still be achieved in the cultivation area despite local non uniformities within the bundle.

A larger axial displacement, Z, would have produced a more uniform flux distribution on the receiver plane. However, this solution would also require larger receiver areas and a higher number of optical fibers, resulting in a significant increase in the initial investment cost. The selected receiver widths therefore represent a trade-off between optical collection efficiency, flux uniformity, and system cost.

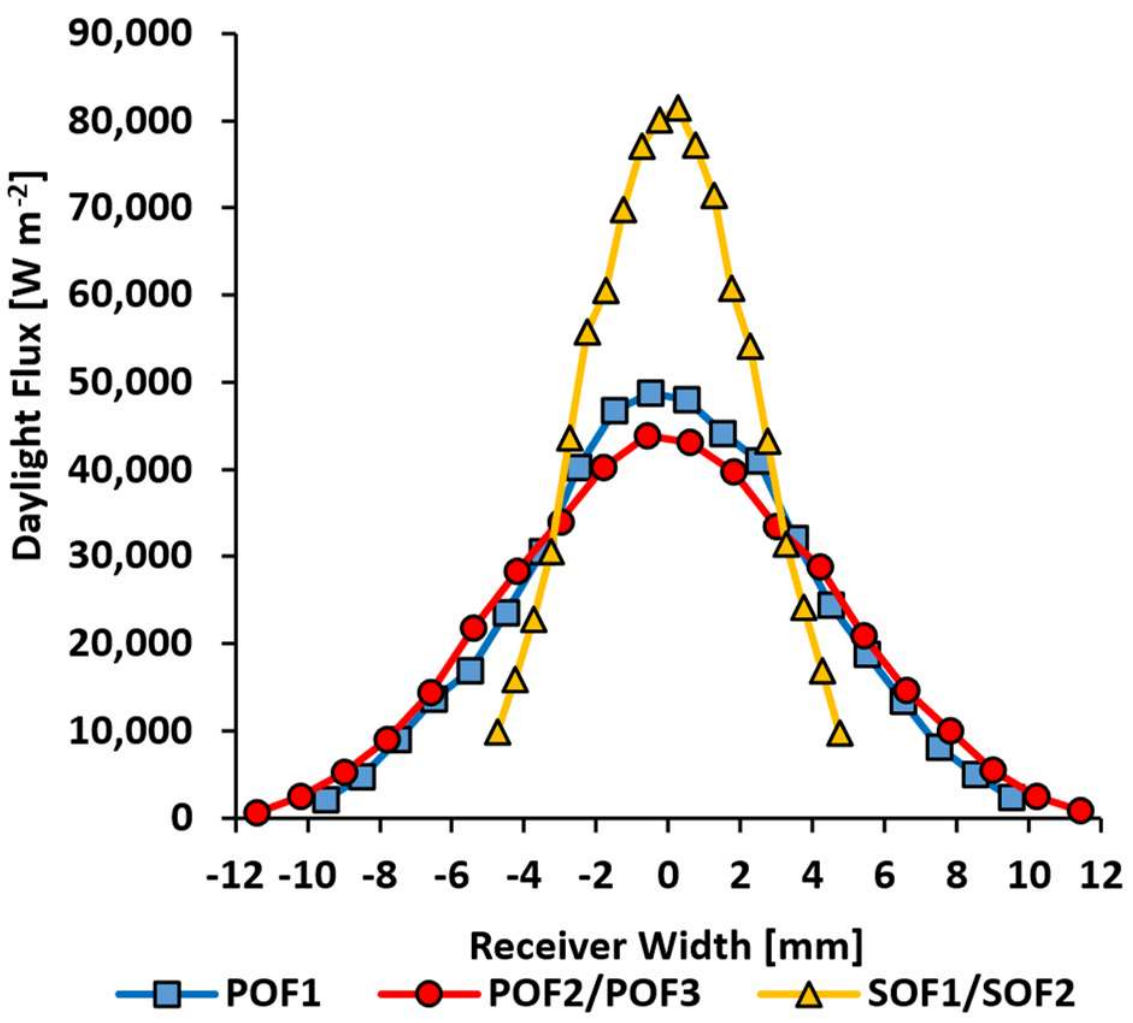

Figure 6: Daylight flux distribution across the receiver width for the selected POF1, POF2/POF3, and SOF1/SOF2 configurations at their optimal axial positions.

To determine the number of collectors that can be installed on the vertical farm rooftop, the spacing between adjacent collectors was assessed. This parameter represents a trade-off between mutual shading, which reduces the amount of daylight intercepted by the Fresnel concentrators, and roof area use, which determines the number of collectors that can be installed within the available surface.

A sensitivity analysis was therefore performed by varying the collector spacing, as shown in Figure 7~~Figure 7~~. For brevity, only the POF1 configuration is reported, since the other configurations exhibited similar trends. As expected, increasing the collector spacing reduces mutual shading, leading to a gradual improvement in shading efficiency. However, this benefit is counterbalanced by a reduction in roof area use, since fewer collectors can be installed on the rooftop.

The overall light capture efficiency, defined as the fraction of the total incident daylight effectively collected by the system, follows the trend of roof area use more closely than that of shading efficiency. This indicates that, within the investigated range, the reduction in the number of installable collectors has a stronger impact on the harvested energy than the improvement obtained by decreasing mutual shading. This behavior is particularly relevant for the considered location, Dubai, where the average solar altitude is relatively high and the largest share of the available solar energy occurs when the Sun is high in the sky. Under these conditions, mutual shading is intrinsically limited, and its effect on the collected energy is less significant than the reduction in the number of collectors caused by increasing the spacing. Consequently, shorter collector spacings are preferable, despite the associated decrease in shading efficiency.

The minimum allowable spacing was ultimately constrained by the system architecture, particularly by the distance required between the Fresnel lens and the receiver. Therefore, the selected collector spacing represents a compromise between optical performance, available rooftop area, and geometrical constraints of the concentration system.

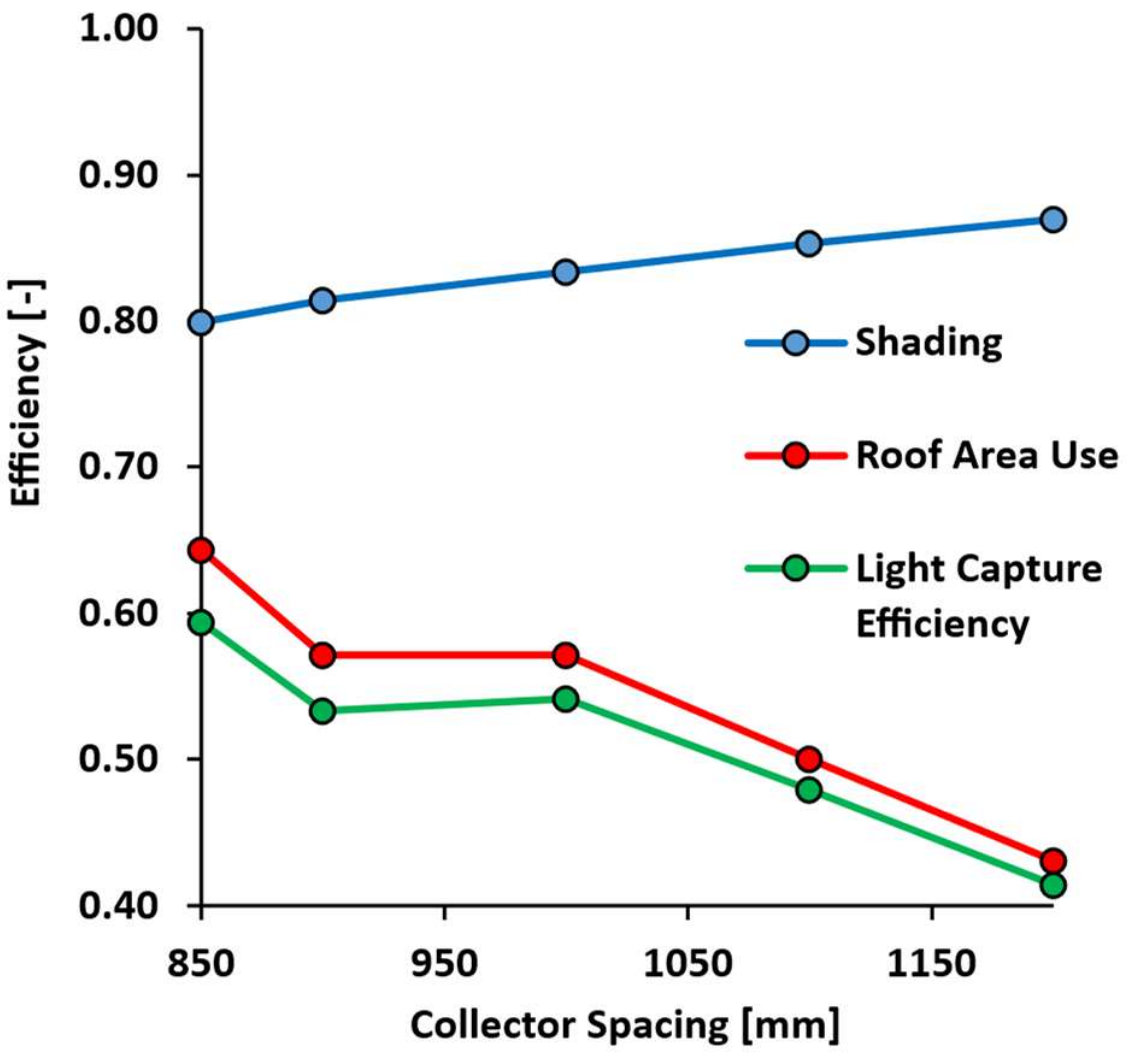

Figure 7: Effect of collector spacing on shading efficiency, roof area use, and overall light capture efficiency for the POF1 configuration.

The construction parameters associated with the five optical-fiber configurations under the MAX strategy are summarized in Table 3~~Table 3~~, whereas those for the MIN strategy are reported in Table A4. Depending on the routing path to the growing zones, the average fiber length is estimated to range from 3 to 6 m. A detailed engineering assessment of routing, connectors, installation, maintenance, and tracking is beyond the scope of this study. The very large number of fibers required by some configurations represents both an economic limitation and a potential practical constraint. In this respect, POF3 reduces the number of individual fibers, while the MIN strategy reduces both the total number of bundles and fibers. Since fiber purchase cost alone already makes the system economically unattractive, excluding additional installation and balance-of-system costs represents an optimistic assumption that does not affect the main conclusion.

Table 3: Geometrical and layout parameters of the Fresnel concentrator and optical fiber receiver for each investigated fiber typology.

| **Parameter** | **POF1** | **POF2** | **POF3** | **SOF1** | **SOF2** | **Unit** |
|---|---|---|---|---|---|---|
| Focal Length | 800.0 | 450.0 | 450.0 | 650.0 | 650.0 | mm |
| Axial Displacement | 15.0 | 10.0 | 10.0 | 0.0 | 0.0 | mm |
| Receiver Width | 20.0 | 24.0 | 24.0 | 10.0 | 10.0 | mm |
| Collector Spacing | 850.0 | 500.0 | 500.0 | 700.0 | 700.0 | mm |
| Collector Length | 7,000 | 6,500 | 6,500 | 6,750 | 6,750 | mm |
| Collector Number | 9 | 14 | 14 | 10 | 10 | Unit |
| Bundle Number | 3,150 | 3,791 | 3,791 | 6,750 | 6,750 | Unit |
| OF Number | 1,452,150 | 2,251,015 | 276,743 | 776,250 | 189,000 | Unit |

### 4.2. Daylight Harvest and System Efficiency

The amount of daylight harvested by the system is a key parameter for estimating the potential electricity savings associated with the reduction of artificial lighting demand. This quantity results from the combined effect of the optical transmission chain and the geometrical layout of the system, including the installable collecting area, receiver size, concentrator spacing, and number of optical fiber bundles.

This interaction is particularly evident under the MAX strategy. As shown in Figure 8Figure 8, SOF-based configurations collect only slightly more daylight than POF2 and POF3, with an increase of 6.4% compared with POF2/POF3 and 26.7% compared with POF1. Although SOFs provide higher optical transmission efficiency, their layout is constrained by the focal length and receiver geometry, which affect the spacing between adjacent collectors and the total collecting area that can be installed on the rooftop. In the optimized MAX configurations, POF2 and POF3 exploit a useful collecting area 35% larger than SOF1 and SOF2, together with a significantly higher number of optical fibers. This larger collecting area compensates for their lower optical transmission efficiency, resulting in a harvested daylight amount comparable to that obtained with SOF-based configurations.

The role of the system layout becomes even more relevant under the MIN strategy, where the number of installed bundles is strongly reduced by assigning only one bundle to each growing zone. In this case, the collected daylight is no longer mainly driven by the total number of installable collectors, but rather by the light transmission capacity of each individual bundle. This strategy reduces the potential energy savings provided by daylighting, while also decreasing the investment cost, which may improve the economic performance of the system.

Under this configuration, SOF-based solutions are penalized by their smaller bundle size. Although SOFs are characterized by higher optical efficiency and a higher concentration ratio, equal to 50 compared with 25 for POF1 and 24.8 for POF2/POF3, the amount of daylight transmitted through a single SOF bundle is lower. This is because the SOF bundle width is 75% smaller than that of POF1 and 83% smaller than that of POF2/POF3. As a result, under the MIN strategy, SOF configurations harvest approximately 50% less daylight than POF-based configurations.

Overall, these results show that the effectiveness of the proposed daylighting system depends on the interaction between optical efficiency, receiver geometry, collecting surface, and layout constraints. For the configurations analyzed in this study, reducing the system to one bundle per growing zone makes SOF-based solutions less suitable for reducing lighting electricity consumption, despite their higher optical efficiency.

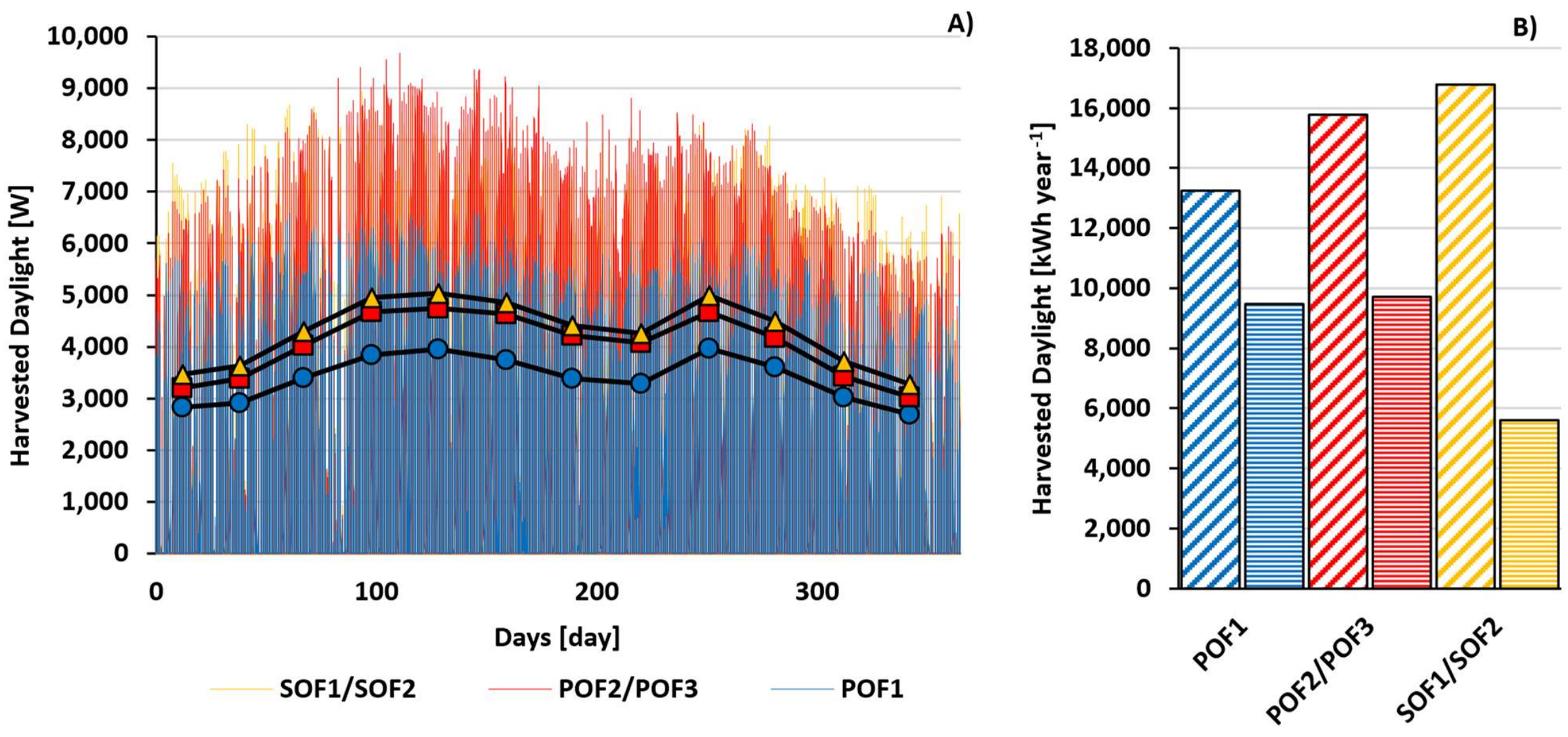

Figure 8: Comparison of daylight harvested by the optical fiber daylighting system. Panel A shows the hourly harvested daylight over the year for the different optical fiber typologies, with markers indicating monthly average values calculated over daylight hours only. Panel B reports the corresponding annual harvested daylight under the MAX and MIN integration strategies. In Panel B, the hatched bars refer to the MAX strategy, whereas the horizontally striped bars refer to the MIN strategy.

The previous considerations are further supported by Figure 9Figure 9, which reports the distribution of the incident sunlight energy among the useful PAR fraction and the main loss contributions occurring along the optical transmission chain. The figure highlights that the final useful PAR energy depends on the combined effect of spectral availability, shading, concentration, coupling, packing, and transmission losses. The simulated optical efficiencies obtained for the POF configurations (18.7-22.6%) are of the same order of magnitude as those reported by Vu et al. [24] , who obtained approximately 19% in simulation and 16% experimentally.

Shading losses are mainly governed by the spacing between adjacent Fresnel concentrators. As discussed above, a shorter collector spacing allows a larger collecting area to be installed on the rooftop, but also increases mutual shading. This effect is particularly evident for the POF2/POF3 configuration, which adopts a smaller spacing and therefore shows higher shading losses than the other configurations. Nevertheless, the larger installable collecting area partly compensates for this penalty in terms of annual harvested daylight.

The concentration and Fresnel-related losses remain relatively limited for all the analyzed configurations, confirming that the selected receiver sizes allow an effective coupling between the Fresnel lens and the optical fiber bundle. Similarly, the losses associated with packing, fiber inlet, bundle distribution, and diffuser transmission remain contained, owing to the square-shaped receiver and the hexagonal packing arrangement of the fibers. This result represents an important requirement for an efficient optical-fiber daylighting system, since excessive coupling losses would strongly reduce the amount of light effectively delivered to the plants.

The main differences among fiber typologies are observed in the transmission losses. These losses are almost negligible for SOF-based configurations, allowing SOF1/SOF2 to reach the highest useful PAR fraction, equal to 26.8%. This value is 43% higher than that obtained for POF2/POF3. However, as shown in the previous analysis, this higher optical performance does not automatically lead to a proportional increase in annual harvested daylight, since the total collected energy is also constrained by receiver size, focal length, collector spacing, and the number of installable bundles.

Overall, the largest reduction is associated with the non-PAR fraction of sunlight, since only part of the incident solar spectrum can be effectively used for crop growth. This contribution does not represent a loss introduced by the optical system, but rather the portion of the solar spectrum that is not useful for crop photosynthesis. Therefore, after excluding this unavoidable spectral fraction, the remaining losses are mainly determined by shading, concentration and coupling effects, and fiber transmission.

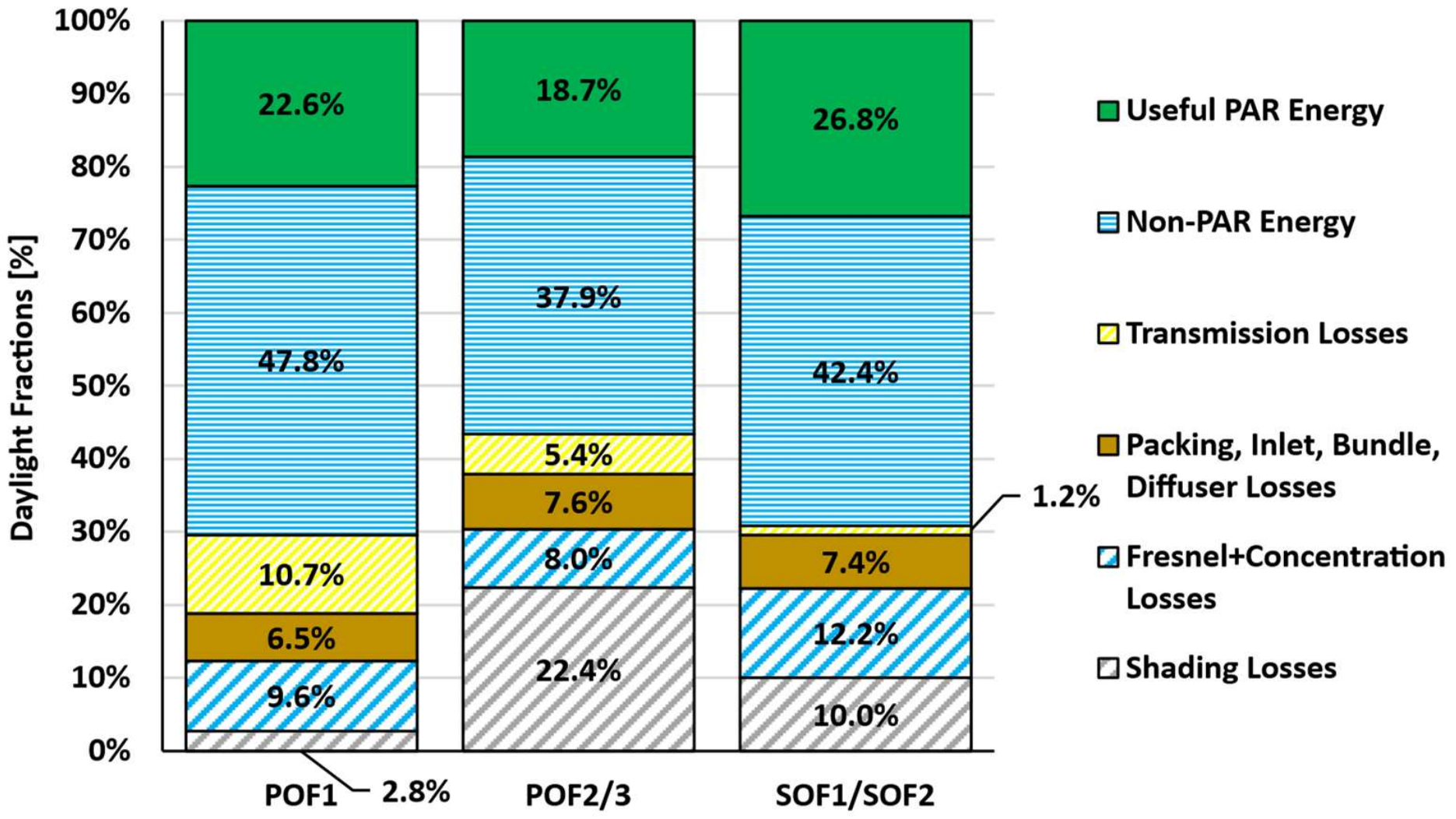

Figure 9: Distribution of the incident daylight energy among useful PAR energy and the main loss contributions along the optical-fiber transmission chain for the analyzed fiber typologies.

### 4.3. Agronomic Standpoint

The PPFD delivered to the crop varies significantly depending on the optical fiber typology and the adopted daylighting strategy. As shown in Figure *10*~~Figure 10~~, the benchmark case maintains a constant PPFD of 250 µmol $m^{-2}$ $s^{-1}$ over the 16 h photoperiod, representing the target lighting condition for crop growth.

In the natural-light-only strategy, referred to as NL, no LED system is used and the light supplied to the plants is entirely provided by the optical-fiber daylighting system. As a result, the delivered PPFD is markedly lower during the morning and evening hours, while it increases around solar noon. During winter, the midday daylight contribution can exceed the benchmark setpoint by approximately 53% for SOF1/SOF2, 28% for POF2/POF3, and 11% for POF1. The use of solar tracking limits the seasonal variation of the peak PPFD, although the total daily light delivered by the system increases during summer due to the longer period of daylight availability.

The relative performance of the different fiber typologies changes with the available solar radiation. In winter, when daylight availability is lower, SOF-based configurations provide higher PPFD due to the greater efficiency of their optical transmission chain. Conversely, in summer, POF2/POF3 configurations compensate for their lower optical efficiency through a larger concentrator area, delivering PPFD values comparable to those obtained with SOF-based configurations.
When the hybrid strategy is operated under the MAX approach, the largest possible number of optical fiber bundles is installed to maximize daylight delivery to the crop. In this case, the LED system compensates for the daylight deficit whenever the optical-fiber contribution is absent or insufficient, allowing the delivered PPFD to remain close to the target setpoint. Deviations from the target occur when the daylight contribution approaches the required PPFD. Due to the minimum dimming level of the PWM-controlled LED driver, equal to 30% of the nominal load, the LEDs cannot be further dimmed once daylight exceeds approximately 175 μmol $m^{-2}$ $s^{-1}$. Under these conditions, the LED system is switched off, producing short periods of mismatch with respect to the target PPFD. These mismatches typically occur at the beginning and end of the day, when the daylight contribution is too high to allow LED operation at the minimum load, but still insufficient to independently meet the target. When daylight exceeds the target PPFD, the LEDs remain off and the delivered PPFD follows the daylight-only profile.
Under the MIN approach, only one optical fiber bundle is assigned to each growing zone, reducing the number of required fibers and, consequently, the initial capital investment. However, this configuration strongly limits the amount of daylight delivered to the crop. None of the analyzed optical fiber typologies is able to meet the target PPFD through daylighting alone, and LED supplementation therefore becomes substantially higher. During the central hours of the day, POF1 and POF2/POF3 can reduce the required LED contribution below the 30% minimum dimming threshold, generating temporary deviations from the desired PPFD setpoint. Conversely, due to the lower daylight collected by SOF1/SOF2 in the MIN configuration, the LED system remains operative throughout the photoperiod and the PPFD setpoint is maintained over the entire year.

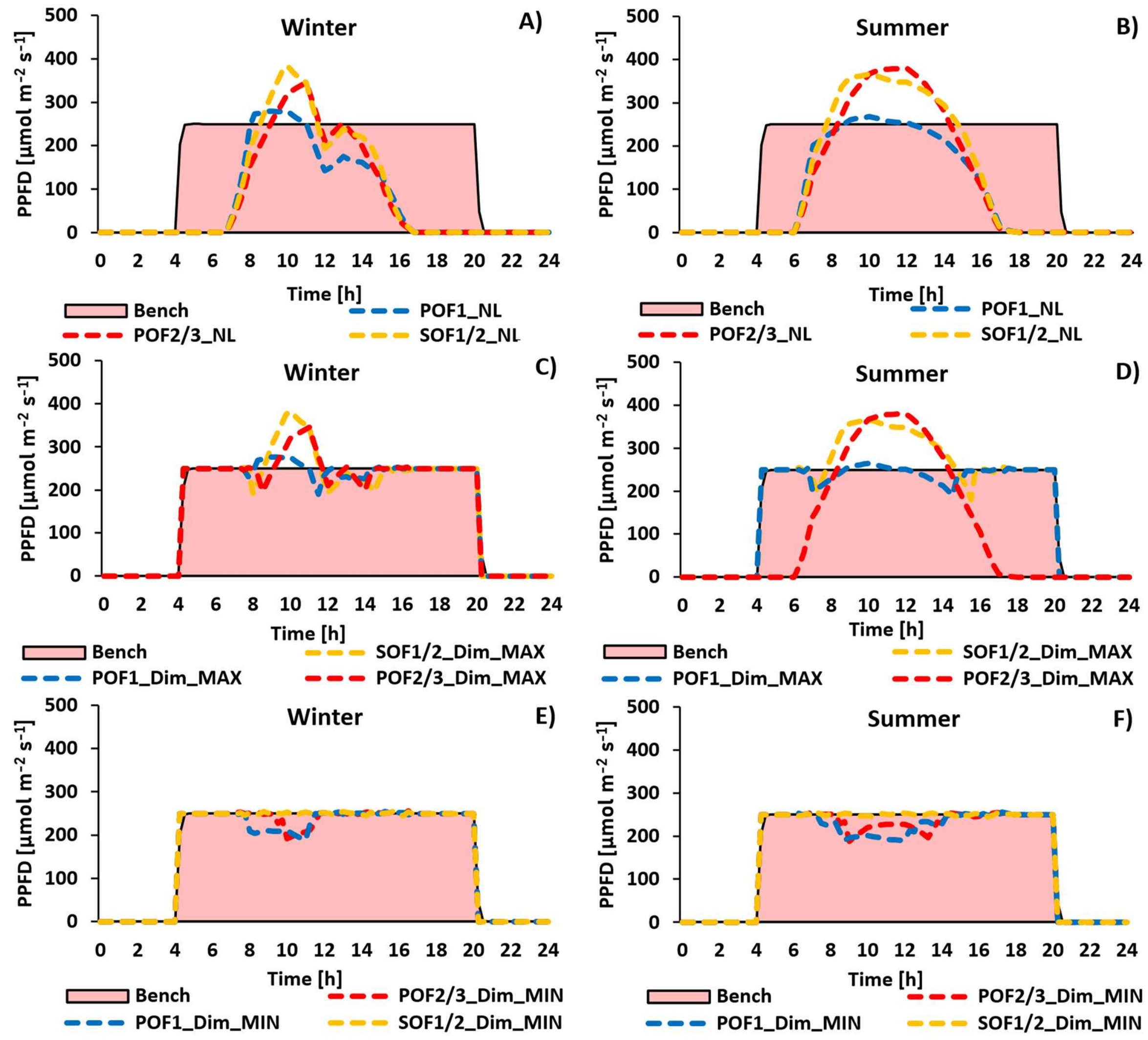

Figure 10: Daily PPFD profiles delivered to the crop under winter and summer conditions for the analyzed optical fiber typologies and daylighting strategies. Panels A and B show the natural-light-only strategy, panels C and D show the hybrid LED–OF strategy under the MAX approach, and panels E-F show the hybrid strategy under the MIN approach. The benchmark profile represents the target PPFD of 250 µmol $m^{-2}$ $s^{-1}$ over the 16 h photoperiod.

### 4.4. Daylighting impact on VF daily power demand

The electricity demand of the vertical farm is directly affected by the performance of the optical-fiber daylighting system. Figure *11*~~Figure 11~~ reports the power demand profiles of the analyzed configurations and compares them with the benchmark scenario, represented by a 16 h photoperiod at a constant PPFD of 250 µmol $m^{-2}$ $s^{-1}$ provided entirely by the LED system. In the benchmark case, the average daily electrical load ranges between approximately 10 and 12 kW, with higher values during summer due to the reduction in the cooling system COP.

The integration of daylighting significantly reduces the lighting load, although the magnitude of this reduction depends on the adopted strategy and optical fiber typology. The additional cooling load associated with daylight introduction remains limited due to the presence of the UV-IR filter, which reduces the transmission of spectral components that would otherwise contribute to heat gains inside the vertical farm.

The NL strategy, where no LED supplementation is used, results in the lowest electricity demand among the investigated scenarios. In this case, the remaining electrical load is mainly associated with cooling requirements, since artificial lighting is completely removed. Under this strategy, the electrical load decreases to approximately 2–4 kW depending on the season, corresponding to a reduction of about 73% compared with the benchmark case.

When LED supplementation is included under the MAX strategy, the electricity demand follows the availability of daylight delivered by the optical-fiber system. During periods without daylight, the load approaches the benchmark value because the target PPFD must be supplied by the LEDs. As the daylight contribution increases, the artificial lighting demand decreases and the total electrical load is consequently reduced. During daylighting hours, the maximum load reaches 6 kW in winter and 4 kW in summer for POF1, 2 kW in winter and 4 kW in summer for POF2/POF3, and 6 kW in winter and 5 kW in summer for SOF1/SOF2. These values remain below the benchmark profile, confirming the effectiveness of the MAX strategy in reducing lighting-related electricity demand.

Under the MAX strategy, the electricity demand profiles of POF2/POF3 and SOF1/SOF2 remain relatively close, despite their different optical properties. As discussed above, this result is related to the compensation between optical efficiency and installable collecting area, which leads to comparable daylight availability at crop level.

The MIN strategy provides lower electricity savings than the MAX strategy, since the number of installed bundles is reduced and LED supplementation is required for a larger portion of the photoperiod. During daylighting hours, the maximum load increases to 7 kW in winter and 8 kW in summer for POF1, 6 kW in winter and 8 kW in summer for POF2/POF3, and 8 kW in winter and 10 kW in summer for SOF1/SOF2. The higher loads observed in summer are mainly related to the lower COP of the cooling system.

Among the MIN configurations, SOF1/SOF2 are the most penalized, since their smaller bundle size limits the daylight delivered to each growing zone. As a result, the optical-fiber contribution remains below the target PPFD throughout most of the daylight period, and the LED system must remain active to maintain the desired lighting condition. Conversely, POF1 is less penalized by the one-bundle-per-growing-zone constraint. Although its optical transmission efficiency is lower than that of SOFs, its larger receiver size allows more daylight to be delivered by each bundle. Compared with POF2/POF3, POF1 also benefits from higher optical transmission efficiency, making it a more balanced configuration under the MIN strategy.

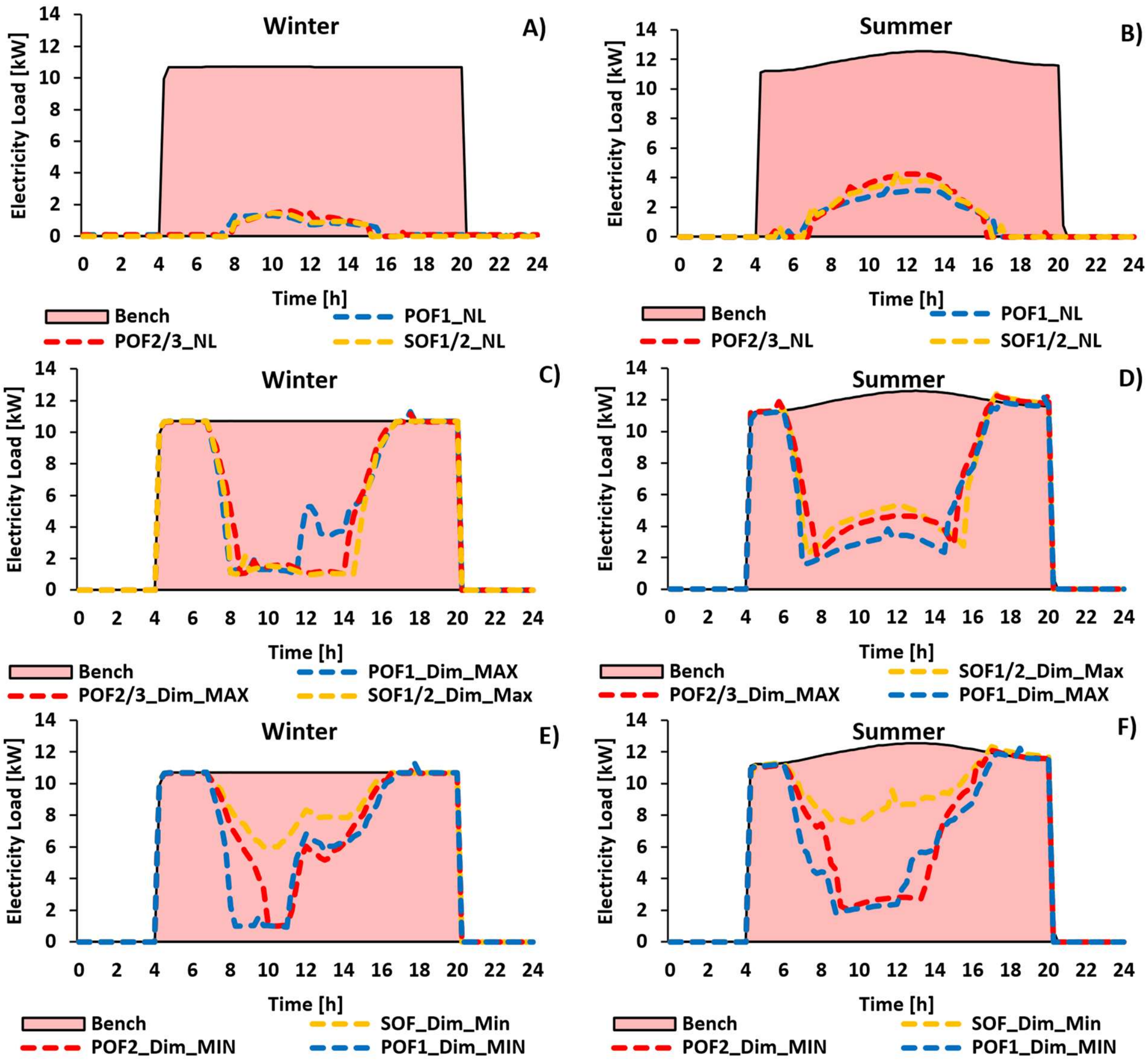

Figure 11: Daily electricity load profiles of the vertical farm under winter and summer conditions for the analyzed optical fiber typologies and daylighting strategies. Panels A and B show the natural-light-only strategy, panels C and D show the hybrid LED–OF strategy under the MAX approach, and panels E and F show the hybrid LED–OF strategy under the MIN approach. The benchmark profile represents the electricity demand of the reference LED-based system operated at a constant PPFD of 250 μmol $m^{-2}$ $s^{-1}$ over the 16 h photoperiod.

### 4.5. Daylighting impact on energy key indicators

From an annual energy-consumption perspective, the benchmark vertical farm is mainly dominated by lighting and cooling loads, due to the continuous use of LED lighting over the 16 h photoperiod. When optical-fiber daylighting is introduced, the lighting demand decreases significantly, with a reduction that depends on the adopted integration strategy and on the amount of daylight delivered to the crop. The use of daylight is associated with an increase in cooling demand, since part of the transmitted solar radiation contributes to the internal heat gains of the vertical farm. However, this increase remains limited due to the presence of the UV-IR filter, which reduces the transmission of spectral components mainly responsible for overheating.

The overall specific energy consumption confirms the energy benefit of the proposed daylighting approach. The benchmark scenario reaches a SEC of 11.2 kWh $kg^{-1}$, while the MAX hybrid strategy reduces this value to approximately 6.6 kWh $kg^{-1}$. Nevertheless, the specific electrical energy consumption is considered the most relevant indicator in this study, since the main objective of the proposed strategy is to reduce the electricity demand associated with artificial lighting and climate control.

In the NL strategy, where no LED supplementation is used, the lighting electricity demand is completely removed. Under this configuration, the cooling load also decreases compared with the benchmark case, since the heat released by the artificial lighting system is avoided. Conversely, the dehumidification demand increases, because the chiller does not directly provide dehumidification under the same operating conditions. Despite this redistribution of energy loads, the absence of artificial lighting allows the NL strategy to achieve the lowest energy use among the analyzed configurations. From an electrical perspective, the NL strategy reaches a SEEC of 1.81 kWh $kg^{-1}$, corresponding to a 76% reduction compared with the benchmark and a 72% reduction compared with the light-pipe-based scenario, as shown in Figure *12*~~Figure 12~~. However, this strong electricity saving is accompanied by a reduction in crop productivity, caused by the lower and less controlled light input delivered to the plants.

In the hybrid MAX strategy, LED supplementation is still required during periods of low daylight availability, leading to higher consumption than the NL case. However, this strategy achieves a substantial reduction in electricity demand while maintaining crop productivity close to, or slightly above, the benchmark level. All MAX optical-fiber configurations achieve an SEEC ranging from 3.16 to 3.37 kWh $kg^{-1}$, corresponding to an approximately 56% reduction compared with the benchmark scenario. This reduction is obtained without relevant productivity losses. In particular, POF2/POF3 and SOF1/SOF2 slightly increase crop production by approximately 3–4% compared with the benchmark case.

When the MIN strategy is applied, the reduced number of optical fiber bundles increases the need for LED supplementation, since less daylight is delivered to the growing zones. As a result, the SEEC increases to 3.78–4.90 kWh $kg^{-1}$, which is higher than the MAX strategy but still below the benchmark value of 7.51 kWh $kg^{-1}$. At the same time, crop productivity remains slightly lower than in the benchmark scenario because the daylight contribution is less effective in maintaining the target lighting conditions. Therefore, the MIN strategy combines lower energy-saving potential with a productivity penalty, making it less attractive from an agro-energy perspective. Its possible advantage is mainly related to the reduction of the initial investment cost, which is assessed in the economic analysis. The energy performance reported in this study is strongly influenced by the high direct solar availability of Dubai. In locations characterized by lower DNI, the daylight contribution delivered through the optical-fiber system would decrease, increasing the need for LED supplementation and reducing the associated electricity savings. Therefore, the present results should be regarded as representative of a favorable climatic scenario for concentrating daylighting systems. Further site-specific analyses would be required to assess their transferability to regions with lower or more variable direct solar availability. The performance of the daylighting system also depends on the ratio between the available rooftop area and the total cultivation area. Increasing the number of cultivation tiers raises the crop lighting demand without increasing the solar collection area, thereby reducing the daylight contribution per unit of growing area and increasing the need for LED supplementation. Therefore, the reported savings results are specific to the roof-to-cultivation-area ratio adopted in the present investigated configuration.

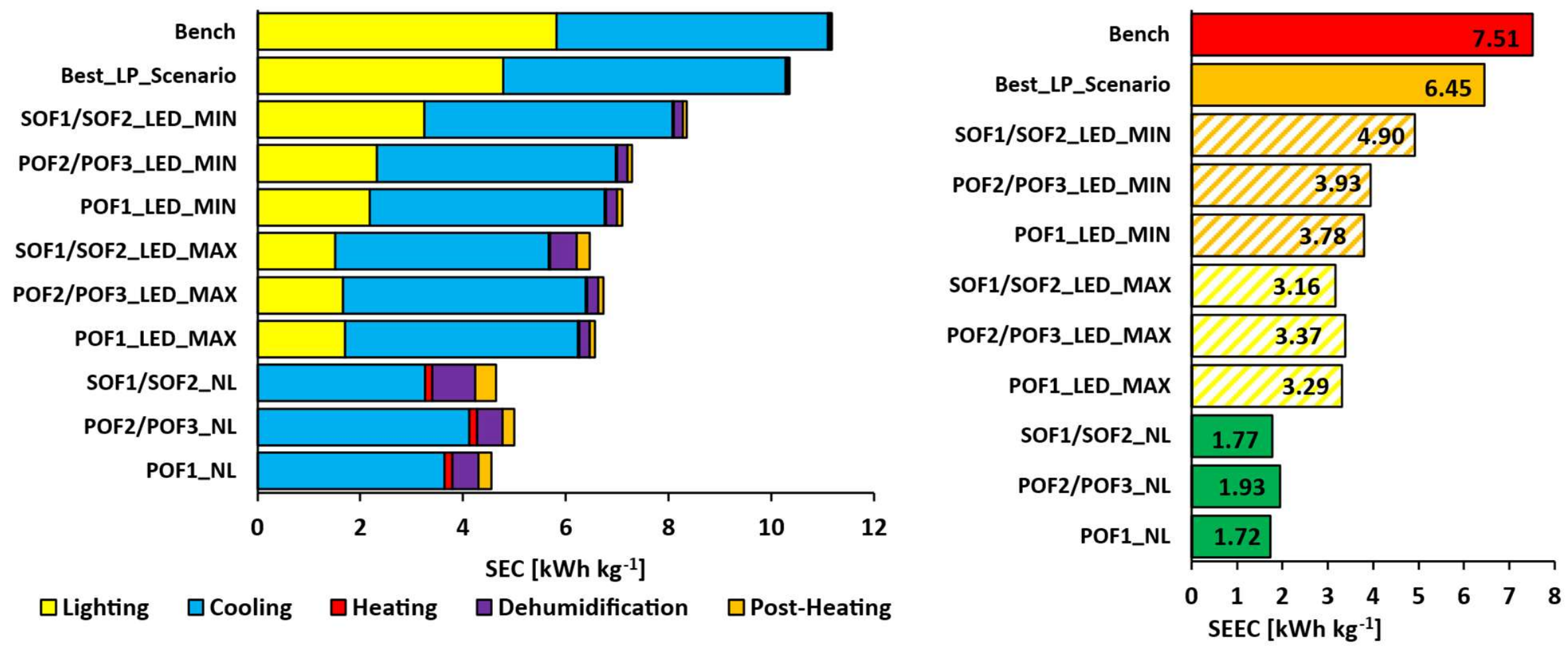

Figure 12: Comparison of the specific energy consumption (SEC) and specific electrical energy consumption (SEEC) of the benchmark, light-pipe reference case, and optical-fiber daylighting configurations. The SEC breakdown includes lighting, cooling, heating, dehumidification, and post-heating contributions, while the SEEC highlights the electrical performance of each strategy.

To complement the SEEC comparison reported in Figure 12 and facilitate a direct comparison among the investigated configurations, Table 4~~Table 4~~ summarizes the annual electricity consumption, crop yield, and the corresponding percentage variations relative to the LED-only benchmark. The table also highlights that electricity savings and SEEC do not vary proportionally, since the latter is normalized by crop productivity.

Table 4: Comparison of annual electricity consumption and crop yield for the investigated optical-fiber configurations. Electricity savings and yield differences are reported relative to the LED-only benchmark.

| **OF** | **Strategy** | **Annual Electricity Consumption [MWh]** | **Yield [kg]** | **Electricity Savings [%]** | **Yield Difference [%]** |
|---|---|---|---|---|---|
| POF1 | NL | 7.5 | 4,380 | 89% | -52% |
| POF1 | MAX | 29.7 | 9,020 | 56% | 0% |
| POF1 | MIN | 33.9 | 8,975 | 50% | -1% |
| POF2/3 | NL | 8.8 | 4,575 | 87% | -50% |
| POF2/3 | MAX | 31.6 | 9,370 | 54% | +3% |
| POF2/3 | MIN | 35.3 | 8,985 | 48% | -1% |
| SOF1/2 | NL | 8.5 | 4,810 | 88% | -47% |
| SOF1/2 | MAX | 29.8 | 9,440 | 56% | +4% |
| SOF1/2 | MIN | 44.4 | 9,060 | 35% | 0% |

### 4.6. Economic Assessment

As discussed in the energy analysis, the optical-fiber daylighting system can achieve higher electricity savings than the light-pipe scenario, mainly because it allows daylight to be distributed across all vertical farm shelves. However, the economic feasibility of this solution remains challenging, since the specific cost of optical fibers is high and the distribution of daylight across the growing area requires a large number of fibers, leading to a high total installed fiber length and a high overall material cost.

The NL strategy was first considered as a theoretical upper limit for electricity savings. However, under the adopted assumptions, this configuration is economically unattractive because the absence of LED supplementation leads to a reduction in crop productivity. Although the NL strategy provides annual electricity savings of approximately 60 MWh, the economic loss associated with the lower crop yield offsets the benefit obtained from reduced electricity consumption. Even assuming an electricity price of 200 $ $MWh^{-1}$, the high selling price of vertically farmed lettuce makes crop productivity the dominant driver of profitability. Therefore, strategies that significantly reduce crop production cannot be considered realistic investment options, even when they provide large electricity savings.

For this reason, the economic analysis was focused on the hybrid strategies, where LED supplementation allows crop productivity to remain close to that of the benchmark scenario. In these cases, the annual electricity savings range from 24 to 38 MWh. The lowest saving is obtained for SOF1/SOF2 under the MIN strategy, while the highest saving is achieved by SOF1/SOF2 under the MAX strategy. These savings are up to 3.87 times higher than those achieved by the light-pipe daylighting system. Nevertheless, the large amount of optical fiber required leads to investment costs that are difficult to recover within the assumed system lifetime of 20 years. Under the current commercial prices considered in this study, none of the analyzed optical-fiber configurations achieves a payback time lower than the expected lifetime. Despite this overall limitation, the economic results provide useful design insights. Although the MAX strategy achieves higher electricity savings, the MIN strategy generally results in a shorter payback time because the reduction in capital expenditure outweighs the associated decrease in annual energy savings. Depending on the optical fiber typology, the MIN strategy reduces the payback time by 9–25% compared with the corresponding MAX configuration. Although the absolute difference between the payback times decreases as the electricity price increases, the MIN strategy maintains the lowest payback time due to its considerably lower capital expenditure. At the same time, higher electricity prices improve the economic attractiveness of the MAX strategy, since its greater electricity savings make the additional investment progressively easier to recover. Therefore, increasing electricity prices reduce the payback-time gap between the two strategies, while the MIN configuration remains the most favorable option under the adopted cost assumptions.

A second important insight concerns the diameter of the optical fibers. Although larger fibers may have higher specific costs and lower packing efficiencies, they strongly reduce the number of fibers required to deliver daylight to the growing zones. Since optical fibers represent the main cost driver of the proposed system, this reduction has a significant impact on the investment cost. In particular, adopting a 3 mm diameter POF reduces the payback time by approximately 60% compared with the corresponding 1 mm diameter POF. Conversely, despite their higher transmission efficiency, SOF-based configurations are less economically attractive under the adopted cost assumptions because of the high specific cost of silica fibers. Under the assumptions of this study, larger POFs combined with short-spacing concentrator layouts represent a more promising solution than SOFs.

To further compare the different daylighting options, a light-cost indicator was evaluated. This indicator was defined as the capital cost required to deliver a unit of photosynthetic photon flux (PPF) to a growing zone under an outdoor daylight availability of 100,000 lux. Figure *13*~~Figure 13~~a reports the photosynthetic photon flux delivered by each solution under this reference condition. Among the optical-fiber configurations, POF2/POF3 provide the highest PPF due to their larger collecting area, with an increase of 47% compared with POF1 and 8% compared with SOF1/SOF2. However, the light-pipe system achieves the highest PPF overall, despite its lower collecting area, because of its shorter and more efficient optical transmission chain.

The corresponding light cost is reported in Figure *13*~~Figure 13~~b. POF1 and SOF-based configurations are not shown because their light cost exceeds 500 $ (µmol $s^{-1}$)$^{-1}$. Despite their higher optical efficiency, the high cost of SOFs makes these configurations economically unattractive under the adopted capital-cost assumptions. Among the optical-fiber solutions, POF3 achieves the lowest light cost, equal to 94 $ (µmol $s^{-1}$)$^{-1}$. This result also supports the use of larger-diameter POFs, which reduce the number of individual fibers required and therefore improve both economic and practical scalability. Nevertheless, this value is almost six times higher than that of the light-pipe solution, confirming that the economic competitiveness of optical-fiber daylighting is currently limited by capital cost rather than by energy performance. The light cost obtained in this study is significantly higher than the value reported for a similar OF-based system in [24] and discussed in [22], equal to 20 $ (µmol $s^{-1}$)$^{-1}$. This difference is mainly related to the lower optical fiber cost assumed in that study. Moreover, the system analyzed in [24] used smaller bundles with fewer optical fibers, which were designed to transport a lower amount of light, approximately 60–75% lower than in the configurations considered here. As a result, the required optical fiber capacity and the corresponding investment cost were considerably lower.

The economic assessment includes optimistic assumptions due to the exclusion of several additional costs, while the use of current commercial component prices does not account for potential economies of scale, which could reduce the cost of optical fibers and other system components. A threshold analysis indicates that, under the adopted assumptions, a cost reduction of approximately 87% would be required for the POF3 configuration to achieve a light cost lower than that of the light-pipe system.

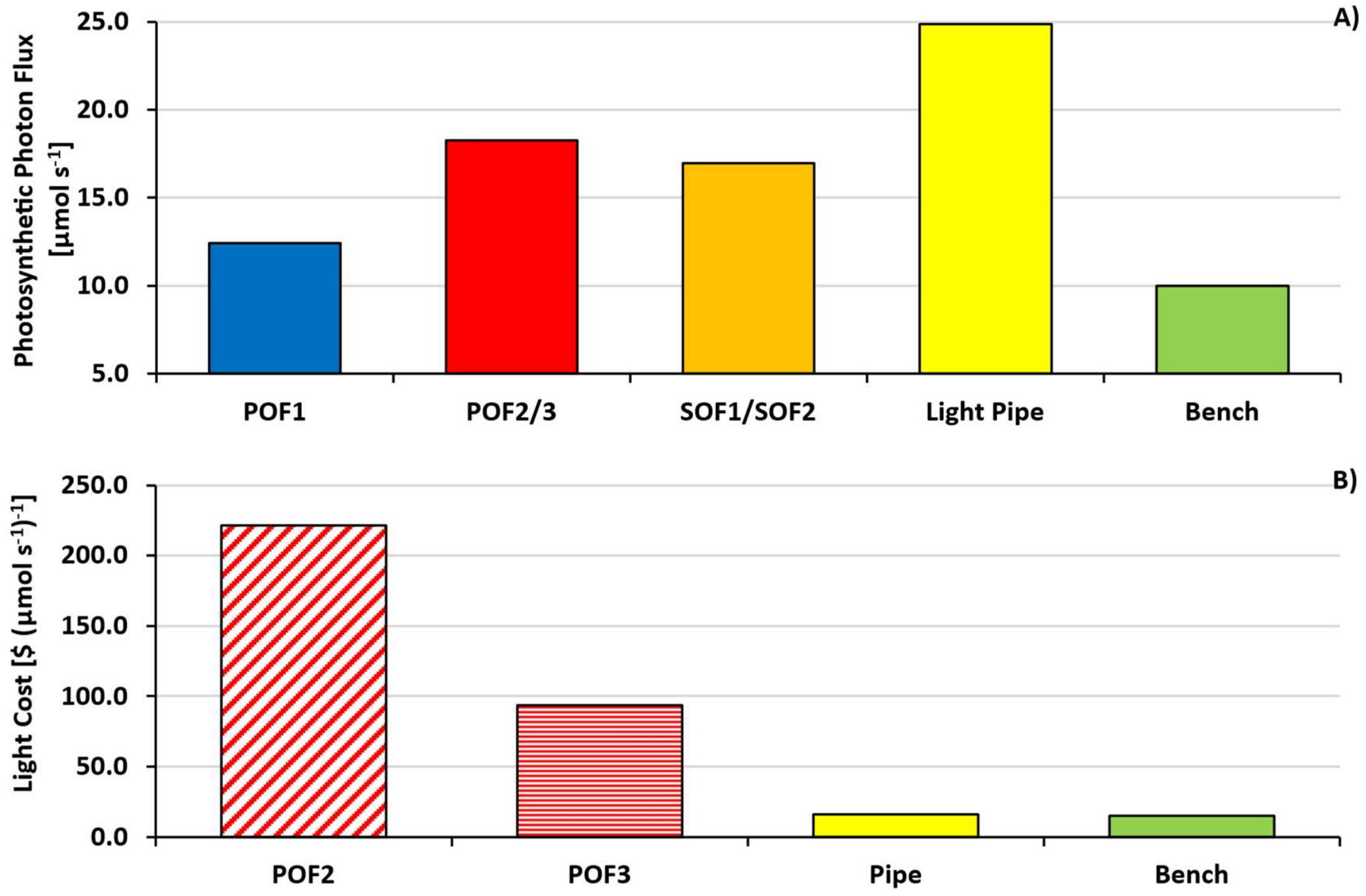

Figure 13: Photosynthetic photon flux delivered to a growing zone under an outdoor daylight availability of 100,000 lux (A) and corresponding light cost for the benchmark, light-pipe, and selected optical-fiber daylighting configurations (B).

To identify the techno-economic parameters that most strongly affect the payback time of the proposed configurations, a one-at-a-time sensitivity analysis was performed. Fiber cost, electricity price, carbon tax, and lettuce selling price were independently varied by ±20%, and their influence was quantified in terms of the percentage variation in PBT relative to the reference case. The reference condition was defined using a POF3 fiber cost of 2.86 \$ m$^{-1}$, electricity price of 0.35 \$ kWh$^{-1}$, carbon tax of 100 \$ t$^{-1}$ and lettuce value of 7.82 \$ kg$^{-1}$. These values were adopted as a common reference for the sensitivity analysis and should not be interpreted as representative market conditions. The analysis was performed for the MIN, MAX, and NL strategies, and the results are reported in Figure 14~~Figure 14~~.

For the hybrid configurations, fiber cost and electricity price are the dominant economic drivers. PBT varies almost linearly with fiber cost, reflecting the dominant contribution of optical fibers to the total system CAPEX. Electricity price also has a substantial effect, resulting in PBT variations of approximately 16–23% for the MIN strategy and 13–18% for the MAX strategy. The influence of carbon tax is comparatively limited under the investigated assumptions. The effect of lettuce selling price is more pronounced when changes in crop productivity contribute substantially to the annual economic balance. This is particularly evident for the NL strategy, where the strong productivity penalty makes lettuce selling price the dominant parameter, with PBT variations ranging from approximately 37% to 142%. Overall, the analysis identifies fiber cost and electricity price as the main economic drivers of the POF3 under MIN strategy, which showed the shortest payback time among the investigated optical-fiber solutions. These two parameters were therefore selected for the subsequent two-parameter sensitivity analysis.

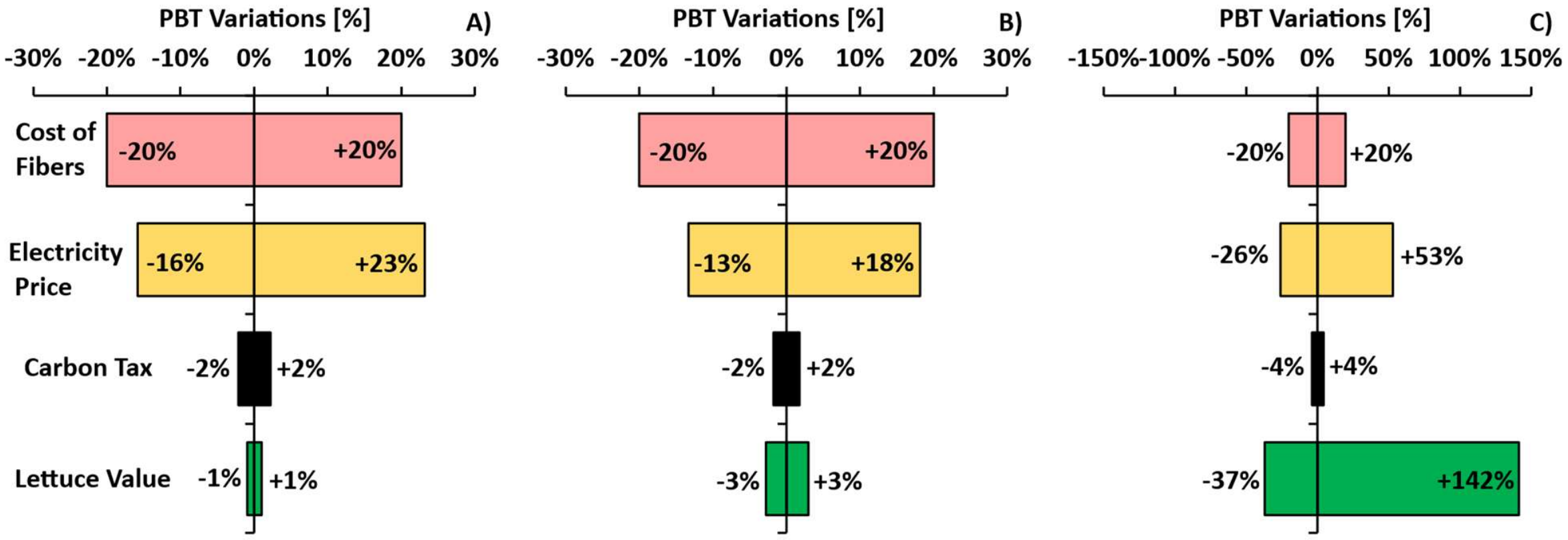

Figure 14: Tornado sensitivity analysis of payback time for the POF3 configuration under the MIN (A), MAX (B), and natural-light-only, NL (C), strategies. Bars show the percentage variation in PBT resulting from an independent ±20% variation in fiber cost, electricity price, carbon tax, and lettuce selling price relative to the corresponding reference case.

Based on these results, a two-parameter payback-time sensitivity analysis was performed for the POF3-MIN configuration, evaluating the combined effects of electricity price and CAPEX reduction, assuming a favorable carbon tax of 100 \$ $t^{-1}$. Based on an emission factor of 404 g $kWh^{-1}$ for the Dubai electricity mix [46], the POF3-MIN configuration avoids approximately 13 $t_{CO2}$ $y^{-1}$. As shown in Figure 15~~Figure 15~~, a CAPEX reduction of about 90%, together with higher electricity prices, is required to bring the payback time within a more acceptable range. Under the assumed conditions, CAPEX reductions below approximately 85% remain insufficient to achieve a payback time shorter than 20 years across most electricity-price scenarios. These results confirm that future developments should primarily focus on reducing optical-fiber costs. The economic feasibility of the system is mainly constrained by the number and purchase cost of the required fibers, while further improvements in optical efficiency would have a comparatively smaller effect on payback time.

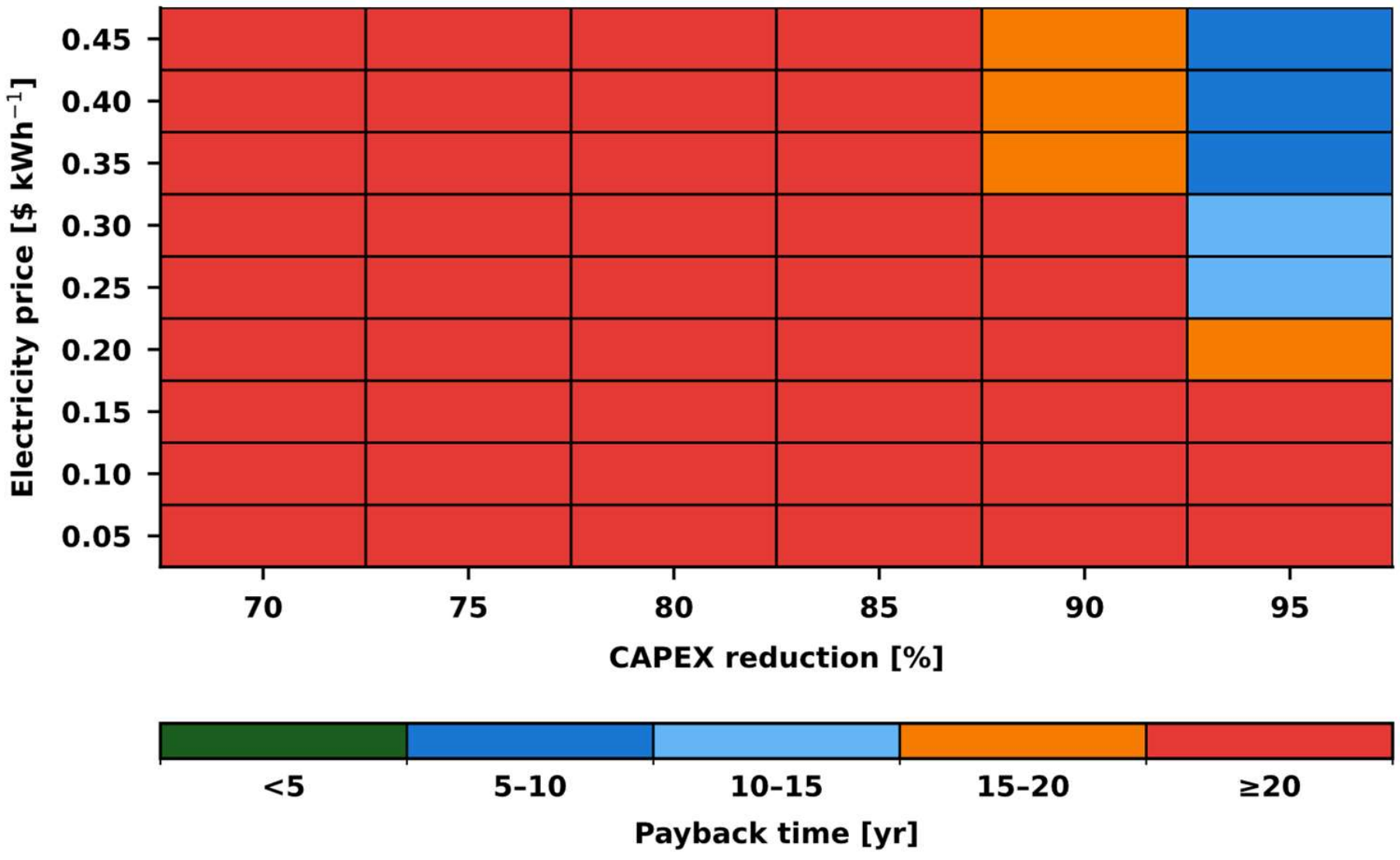

Figure 15: Payback-time sensitivity of the POF3 configuration under the MIN integration strategy as a function of electricity price and optical-fiber system CAPEX reduction.

### 4.7. Alternative Strategy Results and Comparison

To compare the proposed daylighting system with a more conventional solar-based alternative, an additional scenario was evaluated in which no daylighting system was implemented and the available rooftop surface was entirely used for photovoltaic panels. For the selected location, an optimal PV tilt angle of 25° was adopted. A metal supporting frame was considered to install 24 PV panels while avoiding mutual shading between adjacent modules, resulting in an overall installed peak power of approximately 10.5 $kW_{peak}$.

The electricity generated by the PV system was calculated with an hourly time step and compared with the electricity demand of the benchmark vertical farm, as reported in Figure *16*~~Figure 16~~. The benchmark electrical load ranges between approximately 11 and 14 kW depending on the season, while the PV production is limited by the available rooftop area. As a result, the PV system reduces the electricity drawn from the grid, but it does not fully cover the vertical farm electricity demand.

On an annual basis, the PV system produces 15,570 kWh. The hourly simulations show that this production is entirely self-consumed by the vertical farm, covering 22% of its annual electricity demand. By considering the PV electricity production as avoided grid consumption, an equivalent SEEC was calculated by subtracting the generated PV electricity from the benchmark electricity demand. This configuration achieves an equivalent SEEC of 5.79 kWh $kg^{-1}$, which is 47% higher than that obtained with the POF3 configuration operated under the MIN strategy, identified as the most promising optical-fiber case in the economic analysis.

However, despite the lower energy performance of the PV-based solution, its lower capital investment, equal to approximately 30 k$, allows the initial cost to be recovered in about 11 years. This payback time can be considered acceptable for this type of investment and could be further reduced in the presence of incentives, higher electricity prices, or carbon-related economic mechanisms. Furthermore, the relative performance of the two systems may vary with climatic conditions, since the optical-fiber concentrator primarily relies on direct irradiance, whereas the PV system can exploit both direct and diffuse solar radiation.

These results highlight the different maturity levels of the two solar-based strategies. Optical-fiber daylighting provides higher energy-saving potential because it directly reduces the lighting demand inside the vertical farm, but its economic feasibility is currently limited by the high cost of the required optical fibers. Conversely, the PV-based solution provides lower electricity savings, but benefits from much lower capital cost and higher technological maturity. Therefore, for optical-fiber daylighting to become competitive with photovoltaic solutions, future developments should preserve its superior energy performance while substantially reducing the cost of the optical transmission system.

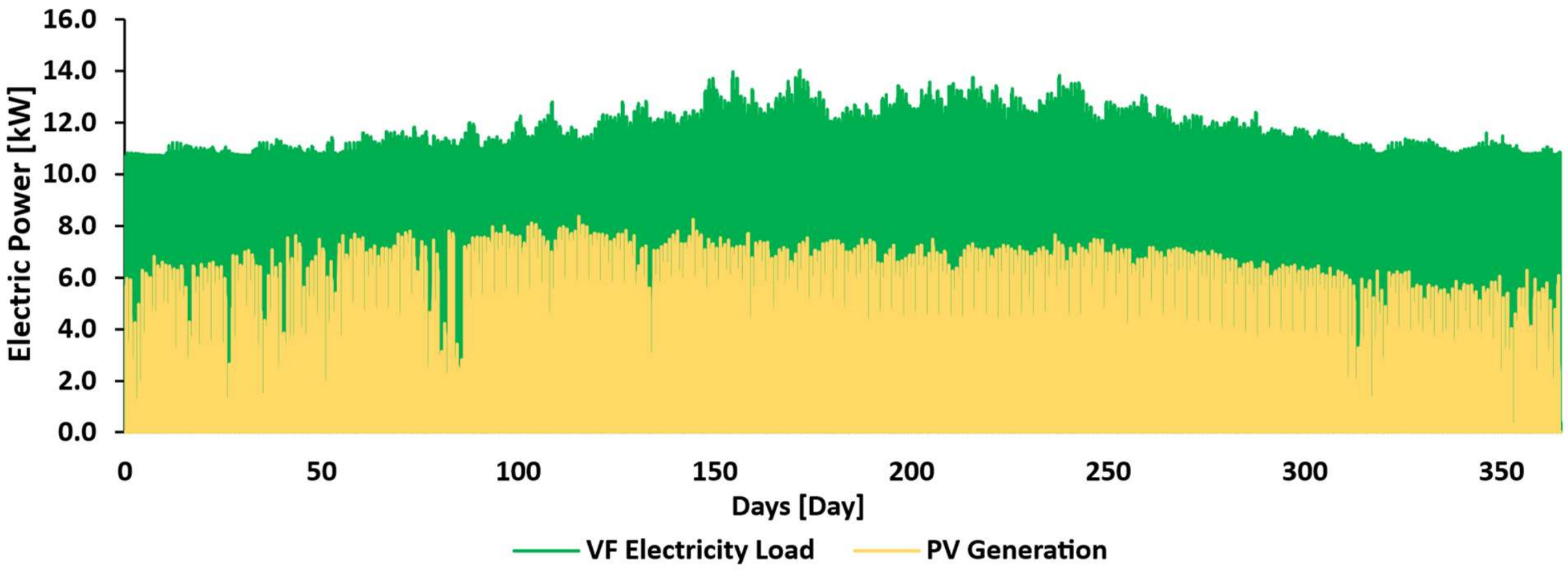

Figure 16: Hourly comparison between the benchmark vertical farm electricity load and the electricity generation from the rooftop PV system over the year.

## 5. Study Limitations

The results of this study should be interpreted in light of some methodological limitations. The analysis is entirely simulation-based and no integrated experimental prototype of the proposed optical-fiber daylighting system was constructed. In addition, the crop-growth model was calibrated using experimental data obtained under controlled lighting conditions, whereas the daylighting configurations investigated here introduce substantially time-varying PPFD. Possible crop responses to short-term fluctuations in light intensity are therefore not explicitly represented and should be investigated experimentally in future work. The economic assessment also relies on a simplified cost framework. The estimated 87–90% reduction in optical-fiber system CAPEX required to approach economic competitiveness does not include costs associated with fiber routing, connectors, installation, tracking electricity consumption, or POF degradation. Including these contributions would further penalize the optical-fiber configurations and would likely increase the required cost reduction.

## 6. Conclusions

This work employed a validated transient model to quantify the techno-economic impact of integrating an innovative optical-fiber daylighting system into a vertical farm. The proposed system was designed to transport concentrated sunlight directly to the crop growing zones, reducing the dependence on artificial lighting. To support the design of the optical-fiber receiver, a single-axis solar-tracked concentration system was modelled in the Tonatiuh ray-tracing software, allowing the optical performance of the concentrator to be assessed while accounting for the operating limits of the selected fibers.

Five optical fiber typologies and three daylighting integration strategies were investigated and compared in terms of agronomic, energy, and economic performance. Crop production, specific electrical energy consumption (SEEC), and light cost (LC) were adopted as the main key performance indicators. The results were also benchmarked against two alternative solar-based strategies: a light-pipe system, characterized by higher optical efficiency but lower integration flexibility, and a photovoltaic-based solution, characterized by lower solar-to-useful-energy performance but higher technological maturity. The main findings and insights are summarized as follows:

- The OF daylighting system achieved an annual useful-PAR delivery efficiency ranging from 19% to 27%, depending on the selected fiber typology. The main limitation is associated with the non-PAR fraction of the solar spectrum, while the most relevant system-related losses are due to the concentration stage and mutual shading between adjacent concentrators.
- The integration of OFs reduced the lighting demand of the vertical farm without substantially increasing the cooling load, mainly due to the use of the UV-IR filter. By distributing daylight across all rack shelves, the system achieved electricity savings ranging from 35% to 89%, depending on the fiber typology and integration strategy, compared with 12% for the light-pipe configuration.
- The daylight-only strategy (NL) achieved the highest electricity saving, with an average SEEC of 1.81 kWh $kg^{-1}$. However, the absence of LED supplementation caused a reduction of up to 52% in crop productivity, making this strategy economically unattractive under all the evaluated scenarios.
- The MAX strategy avoided productivity losses, whereas the MIN strategy substantially reduced the penalty observed in the NL case. The hybrid strategies reduced SEEC by approximately 35–58%, depending on fiber typology and integration approach. The MAX strategy provided the best agro-energy compromise, with no productivity loss and a minimum SEEC of 3.16 kWh $kg^{-1}$ for the high-efficiency SOF configuration.
- The high specific cost of SOFs makes them unsuitable for the proposed system under the adopted economic assumptions. For POFs, the large number of fibers required to distribute daylight throughout the vertical farm also leads to investment costs that are difficult to recover within the system lifetime. A reduction of about 90% in optical fiber capital costs would be required to make the system economically competitive under favorable operating conditions.
- The rooftop PV scenario highlights the economic gap between optical-fiber daylighting and mature solar technologies. Although PV covered only 22% of the annual electricity demand and achieved lower energy savings than the optical-fiber strategies, its lower capital cost allowed a payback time of approximately 11–12 years. Therefore, PV is currently more economically viable, while optical-fiber daylighting remains more effective in directly reducing lighting-related electricity demand.

Based on the findings of this work and on previous studies, daylighting for vertical farms cannot yet be considered a mature and economically feasible solution under current market conditions. High investment costs, uncertainties in light distribution within the farm, and limited technological maturity still offset the benefits associated with electricity savings. Therefore, under the assumptions and market conditions considered in this study, sunlight cannot currently replace LEDs in vertical farms without significant agronomic or economic penalties, although it can substantially reduce their use when integrated through hybrid daylight–LED strategies. Under the same conditions, PV-based systems currently appear to be the more practical and economically favorable option for exploiting the available rooftop solar resource.

## 7. Appendix and supplementary data

### 7.1 Climatic Data

The vertical farm analyzed in this study was assumed to be located in Dubai. To quantify transmission losses and assess natural light availability, climate data for outdoor air temperature, and DNI were collected, and the corresponding profiles are reported in Figure A1. Relative humidity and sky infrared radiation were also included in the analysis, but are not reported here because they play a secondary role in this study.

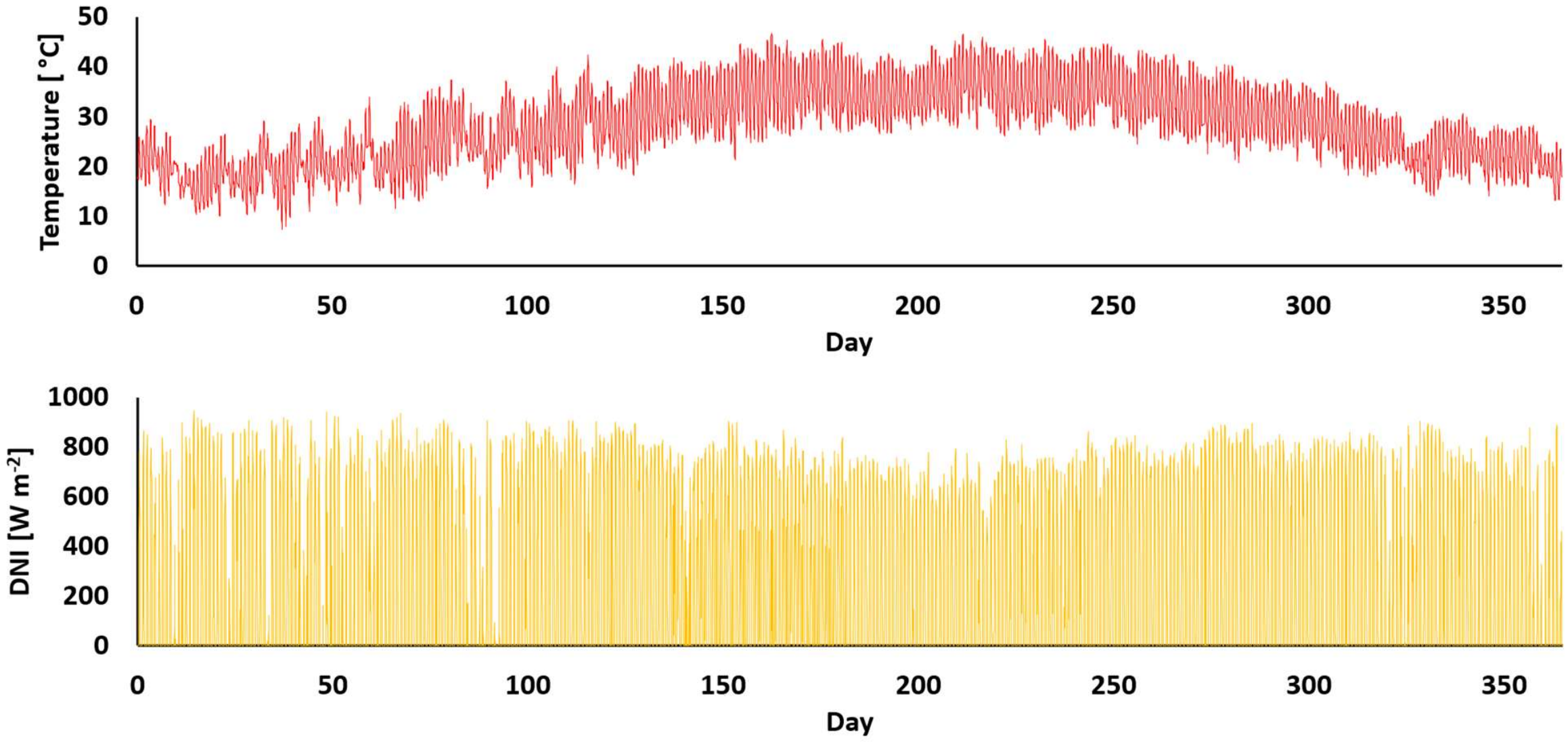

Figure A1: Dubai climate data: outdoor temperature, and DNI.

### 7.2 Solar Position and Angle of Incidence

The solar position is defined by solar altitude ($\alpha_{sol}$) and azimuth ($\gamma_{sol}$), which depend on geographical location and day of the year. Cooper's equation is used to estimate the solar declination δ (Eq. A1), where n is the day of the year. The Equation of Time EoT (Eq. A2) is used to compute the local solar time $h_{sol}$ from clock time $h_{real}$ (Eq. A3). The auxiliary angle B is computed with Eq. A4, and the hour angle ω is obtained from Eq. A5. Solar altitude and azimuth were then computed using Eqs. A6-A7. The site coordinates were set to latitude (φ) 25° and longitude (ψ) 55°. The reference longitude is 60°, 15*ΔUTC for ΔUTC equal to 4. The resulting solar altitude and azimuth values (with North = 0°) are reported in Figure A2.

$$\delta = 23.45 \cdot \sin(360 \cdot \frac{284+}{365}) \tag{A1}$$

$$EoT = 9.87 \cdot \sin 2B - 7.53 \cdot \cos B - 1.5 \cdot \sin B \tag{A2}$$

$$h_{sol} = h_{real} + (EoT - 4 \cdot (\psi - \psi_{ref}))/60 \tag{A3}$$

$$B = (n - 81) * \frac{360}{364} \tag{A4}$$

$$\omega = 15 \cdot (h_{sol} - 12) \tag{A5}$$

$$\alpha_{sol} = \arcsin(\sin\delta \cdot \sin\varphi + \cos\delta \cdot \cos\varphi \cdot \cos\omega) \tag{A6}$$

$$\gamma_{sol} = \arcsin(\frac{\cos\delta \cdot \sin\omega}{co \quad _{sol}}) \tag{A7}$$

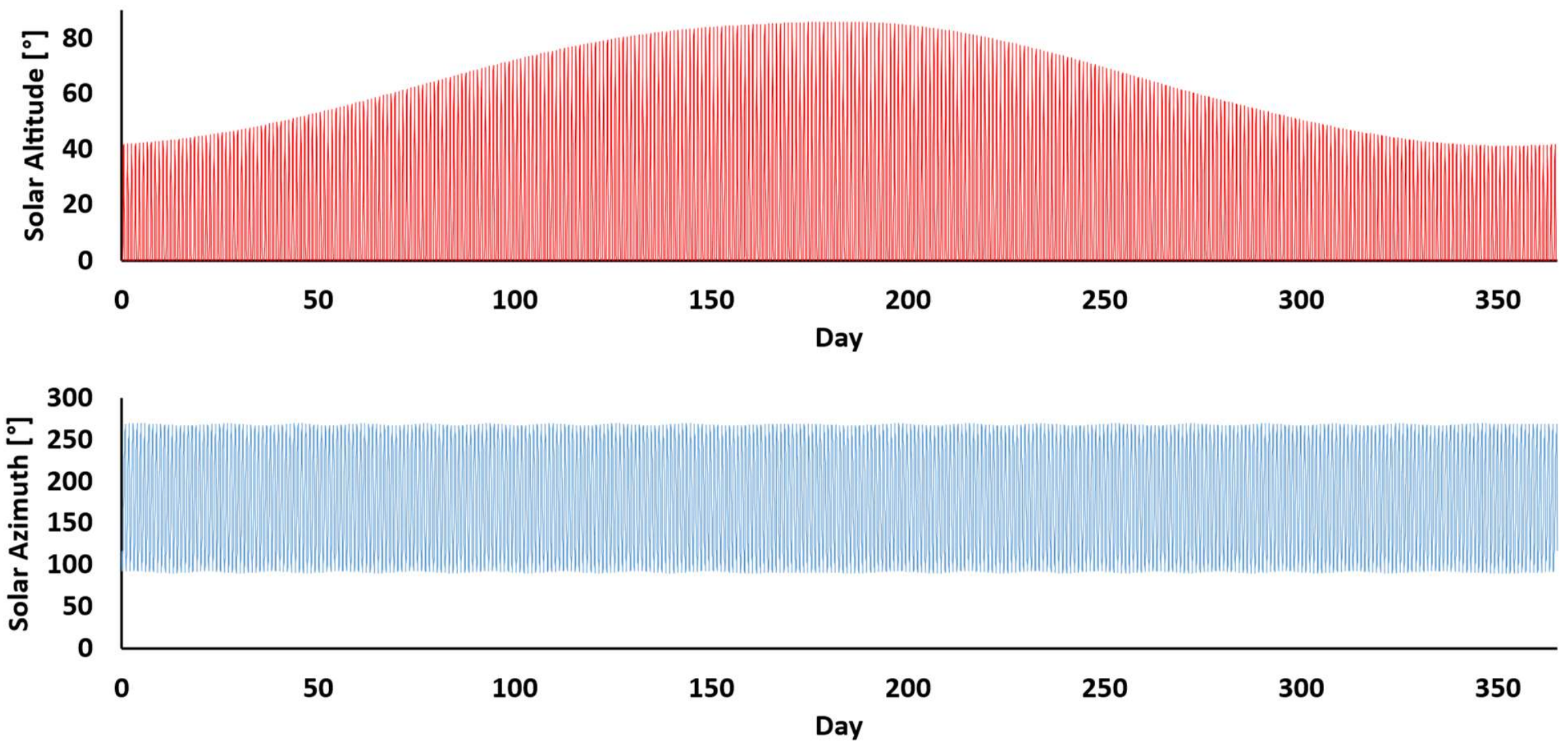

Figure A2: Solar positions in Dubai, including altitude and azimuth angles, referenced to north.

### 7.3 Thermal Model: Assumptions and formulations

This appendix summarizes the assumptions and formulations adopted for the thermal energy balance model used to estimate the equilibrium surface temperature of the optical fiber. The model includes the optical power input, the transmitted and absorbed contributions, and the heat exchange mechanisms associated with conduction, convection, and thermal radiation. The environmental conditions and material properties used in the calculations are reported in Table A1.

Table A1: Assumed thermo-optical properties, environmental boundary conditions, and optical parameters used in the thermal balance model of the optical fiber system. The attenuation coefficient adopted for each fiber was selected as a representative value over the PAR spectral range.

| Name | Symbol | Value | Unit | Reference |
|---|---|---|---|---|
| **OF Diameter** | $D$ | 1.00 | mm | [24] |
| **Cladding Thickness** | $t_{clad}$ | 0.01 | mm | Assumed |
| **OF Length** | $L$ | 5.0 | cm | Assumed |
| **Heat transfer coefficient** | $h$ | 25 | W m$^{-2}$ K$^{-1}$ | Typical Value |
| **Core Emissivity** | $\varepsilon_{core}$ | 0.92 | [-] | [47] |
| **Cladding Emissivity** | $\varepsilon_{clad}$ | 0.90 | [-] | Assumed |
| **Filler Emissivity** | $\varepsilon_{filler}$ | 0.92 | [-] | [48] |
| **Core Attenuation** | $\varsigma_{core}$ | 0.25 | dB m$^{-1}$ | [24] |
| **Cladding Attenuation** | $\varsigma_{clad}$ | 0.87 | dB m$^{-1}$ | Assumed |
| **Filler Attenuation** | $\varsigma_{filler}$ | 980 | dB m$^{-1}$ | Assumed |
| **Cladding Thermal Conductivity** | $\lambda_{clad}$ | 0.12 | W m$^{-1}$ K$^{-1}$ | [49] |
| **Filler Thermal Conductivity** | $\lambda_{filler}$ | 0.19 | W m$^{-1}$ K$^{-1}$ | [50] |
| **Outdoor Temperature** | $T_{out}$ | 40 | °C | [32] |
| **Sky Temperature** | $T_{sky}$ | 27 | °C | [32] |
| **Fresnel Transmissivity** | $\tau_{Fresnel}$ | 0.92 | [-] | [24] |
| **UV-IR Transmissivity** | $\tau_{UV\text{-}IR}$ | 0.45 | [-] | [24] |
| **Concentration Effectiveness** | $\eta_{conc}$ | 0.98 | [-] | Tonatiuh |

To model the filler as an equivalent cylindrical layer, its cross-sectional area in the hexagonal packing configuration was first evaluated. The reference area was defined as the hexagon whose vertices correspond to the centers of the six optical fibers surrounding the central one. The filler area was then obtained by subtracting from this hexagonal area the portions occupied by the optical fibers, as expressed in Eq. (A9).
The resulting filler area was used to define an equivalent annular cylindrical layer surrounding the cladding. The thickness of this equivalent filler layer was calculated by imposing the equality between the actual filler area in the hexagonal packing and the area of the equivalent annulus, as reported in Eq. (A10).

$$A_{filler} = \frac{3\sqrt{3}}{2} \cdot (\frac{D}{2})^2 - \frac{\pi D^2}{4} - 6 \cdot \frac{\pi D^2}{4} \cdot \frac{1}{3} \tag{A9}$$

$$A_{filler} = \frac{\pi (D + t_{filler})^2}{4} - \frac{\pi D^2}{4} \tag{A10}$$

The equivalent cylindrical representation was also used to evaluate the conductive thermal resistances between the different material regions. In particular, the core–cladding and cladding–filler interfaces were modelled as concentric cylindrical layers, allowing the radial conductive heat transfer to be described through the thermal resistance formulation for multilayer cylinders. Based on this approximation, the conductive resistances associated with the core–cladding and cladding–filler heat exchange were calculated as reported in Eqs. (A11) and (A12), respectively.

$$R_{cond,\text{core}-clad} = \frac{\ln \frac{\frac{D}{2} - 0.5 \cdot t_{clad}}{\frac{D}{2} - t_{clad}}}{2\pi L \lambda_{clad}} \tag{A11}$$

$$R_{cond,\text{clad}-fille} = \frac{\ln \frac{\frac{D}{2}}{\frac{D}{2} - 0.5 \cdot t_{clad}}}{2\pi L \lambda_{clad}} + \frac{\ln \frac{\frac{D}{2} + 0.5 \cdot t_{filler}}{\frac{D}{2}}}{2\pi L \lambda_{filler}} \tag{A12}$$

Finally, Eqs. (A13)–(A19) were used to model the individual terms included in the thermal energy balance of the optical fiber. These equations define the contributions associated with the incoming concentrated power, optical transmission, absorption, and heat exchange by conduction, convection, and thermal radiation. The subscript (i) was introduced to indicate the generic material region considered in the calculation. Therefore, the same formulation was applied to the core, cladding, and filler layers.

$$Q_{in,i} = Q_{conc} \cdot A_i \cdot \tau_{Fresnel} \cdot \tau_{UV-I} \cdot \eta_{conc} \tag{A13}$$

$$Q_{tr,i} = Q_{in,i} \cdot 10^{-\frac{\varsigma_i \cdot L}{10}} \tag{A14}$$

$$Q_{cond,core-cl} = \frac{T_{core} - T_{clad}}{R_{cond,core-clad}} \tag{A15}$$

$$Q_{cond,clad-fill} = \frac{T_{clad} - T_{filler}}{R_{cond,clad-fill}} \tag{A16}$$

$$Q_{cond,filler-ext} = 2 \cdot Q_{cond,clad-filler} \tag{A17}$$

$$Q_{conv,i} = hA_i \cdot (T_i - T_{ext}) \tag{A18}$$

$$Q_{rad,i} = \sigma A_i \varepsilon_i \cdot (T_i^4 - T_{sky}^4) \tag{A19}$$

**7.4 Lens Efficiency: Assumptions and Formulations to account for mutual shading**

The geometrical relationships used to evaluate mutual shading between adjacent Fresnel concentrators are reported in this appendix. The formulation was applied at each hourly time step, using the instantaneous collector tilt angle obtained from the monoaxial tracking strategy.

First, the apparent height of the tilted concentrator was calculated using Eq. (A20). This parameter represents the vertical projection of the concentrating module and depends on the collector geometry, the focal length, and the instantaneous tilt angle. It defines the obstruction generated by each concentrator with respect to the adjacent unit.

$$H = W \cdot \sin \alpha_{Fresnel} \quad \text{(A20)}$$

Starting from the apparent height, Eq. (A22) was used to determine the minimum edge-to-edge distance required to avoid mutual shading between two consecutive concentrators. This distance represents the spacing that ensures full illumination of the downstream lens for the considered solar position.

$$\alpha_{sol,corr} = \tan^{-1} \frac{\tan \alpha_{sol}}{\cos(\gamma_{sol} - \gamma_{Fresnel})} \quad \text{(A21)}$$

$$\Delta x_{edges,min} = H \cdot \cot \alpha_{sol,corr} \quad \text{(A22)}$$

The minimum edge-to-edge distance was then converted into the corresponding ideal center-to-center spacing by means of Eq. (A23). This conversion accounts for the physical width of the concentrator and provides the reference spacing between adjacent collector axes under shading-free conditions.

$$\Delta x_{centers,min} = W \cdot \cos \alpha_{Fresnel} + \Delta x_{edges,min} \quad \text{(A23)}$$

For each assumed center-to-center spacing, the difference between the selected spacing and the ideal shading-free spacing was evaluated. This difference was used to identify the occurrence and extent of mutual shading. When the selected spacing is lower than the ideal spacing, the upstream concentrator partially shades the following lens.

The shaded portion of the lens aperture was then calculated using Eqs. (A24). This equation quantifies the fraction of the lens width that is not directly illuminated due to the shadow cast by the preceding concentrator. The remaining illuminated aperture was consequently used to compute the effective collecting area of each lens.

$$W_{shaded} = \left(\Delta x_{centers,min} - \Delta x_{centers}\right) \cdot \frac{\tan\ _{sol,corr}}{(\tan \alpha_{sol,corr} + \tan \alpha_{Fresnel}) \cdot \cos \alpha_{Fresnel}} \quad \text{(A24)}$$

Finally, Eq. (A25) was used to evaluate the overall lens efficiency under the selected spacing configuration. This efficiency was defined as the ratio between the direct solar energy effectively collected by the concentrator array, after accounting for mutual shading and the number of installed units, and the direct solar energy intercepted by an equivalent unobstructed collecting surface.

$$\eta_{shade} = \frac{1 + (N_{Fresnel} - 1) \cdot (1 - \frac{W_{shaded}}{W})}{N_{Fresnel}} \quad \text{(A25)}$$

The evaluation of the optical transmission chain was completed by accounting for the losses occurring at the fiber inlet and during light propagation along the optical fiber. The coupling losses associated with the light entering the fiber bundle were calculated using Eq. (A26), while the transmission losses along the fiber length were evaluated according to Eq. (A27). The packaging efficiency, $\eta_{pack}$, was determined from the hexagonal packing geometry and the fraction of the fiber cross-section occupied by the optical core, as expressed in Eq. (A28), resulting in a value of 0.87. The bundle efficiency, $\eta_{bundle}$, which accounts for additional bundle-related losses, ranged from 0.95 to 0.99 depending on the considered configuration.

$$\eta_{inlet} = 1 - \left(\frac{n_{core} - 1}{n_{core} + 1}\right)^2 \quad \text{(A26)}$$

$$\eta_{tr} = 10^{-\frac{\varsigma \cdot L}{10}} \quad \text{(A27)}$$

$$\eta_{pack} = \frac{\pi}{2\sqrt{3}} \cdot \left(\frac{D - 2t_{clad}}{D}\right)^2 \quad \text{(A28)}$$

**7.5 PV Panels: Assumptions and Formulations to estimate electrical efficiency**

The assumptions adopted for the PV energy model are summarized in this appendix. The reference PV module was selected from [34], which reports the main technical characteristics of the panel, including its dimensions, peak power, nominal efficiency, and the dependence of the module efficiency on the operating temperature. The direct and diffuse irradiance components incident on the tilted PV surface were evaluated separately. Reflection losses for both direct and diffuse radiation were calculated using the model proposed in [35], adopting the correction coefficients reported in that study for monocrystalline silicon modules (see Eqs. (A29-A30).

To account for the effect of operating temperature on the nominal conversion efficiency of the PV module, the cell temperature was estimated according to the approach presented in [36]. The resulting cell temperature was then used to correct the nominal module efficiency under actual operating conditions. The inverter efficiency was evaluated as a function of the power load ratio by using the commercial performance data reported in [37] and summarized in Figure A3. Additional system-level losses, including DC wiring losses, mismatch losses, connection losses, LID losses, nameplate rating, and availability, were grouped into the balance-of-system efficiency, $\eta_{BOS}$, according to the assumptions reported in [38]. Finally, the optimal tilt angle of the PV panels for the selected location was identified using the tools available in PVGIS [32].

$$\rho_{dir} = 1 - \frac{e^{-\frac{\cos\ _{sol}}{a_r}} - e^{-\frac{1}{a_r}}}{1 - e^{-\frac{1}{a_r}}} \tag{A29}$$

$$\rho_{diff} = 1 - e^{-\frac{1}{a_r}\cdot\left(c_1\cdot\left(\sin\ _{PV} + \frac{\pi - \alpha_{PV} - \sin\alpha_{PV}}{1 + \cos\ _{PV}}\right) + c_2\cdot\left(\sin\alpha_{PV} + \frac{\pi - \alpha_{PV} - \sin\alpha_{PV}}{1 + \cos\alpha_{PV}}\right)^2\right)} \tag{A30}$$

$$\eta_T = \eta_{PV} \cdot (1 - \beta \cdot (T_{PV} - T_{ref})) \tag{A31}$$

$$T_{PV} = T_{ext} + \left(I_{dir}\cos\theta + I_{diff}\cdot\left(\frac{1 + \cos\alpha_{PV}}{2}\right)\right)\cdot\omega\cdot\frac{0.32}{8.91 + 2.0\cdot v_{PV}} \tag{A32}$$

$$v_{PV} = \frac{v_{wind} + 0.5}{0.68} \tag{A33}$$

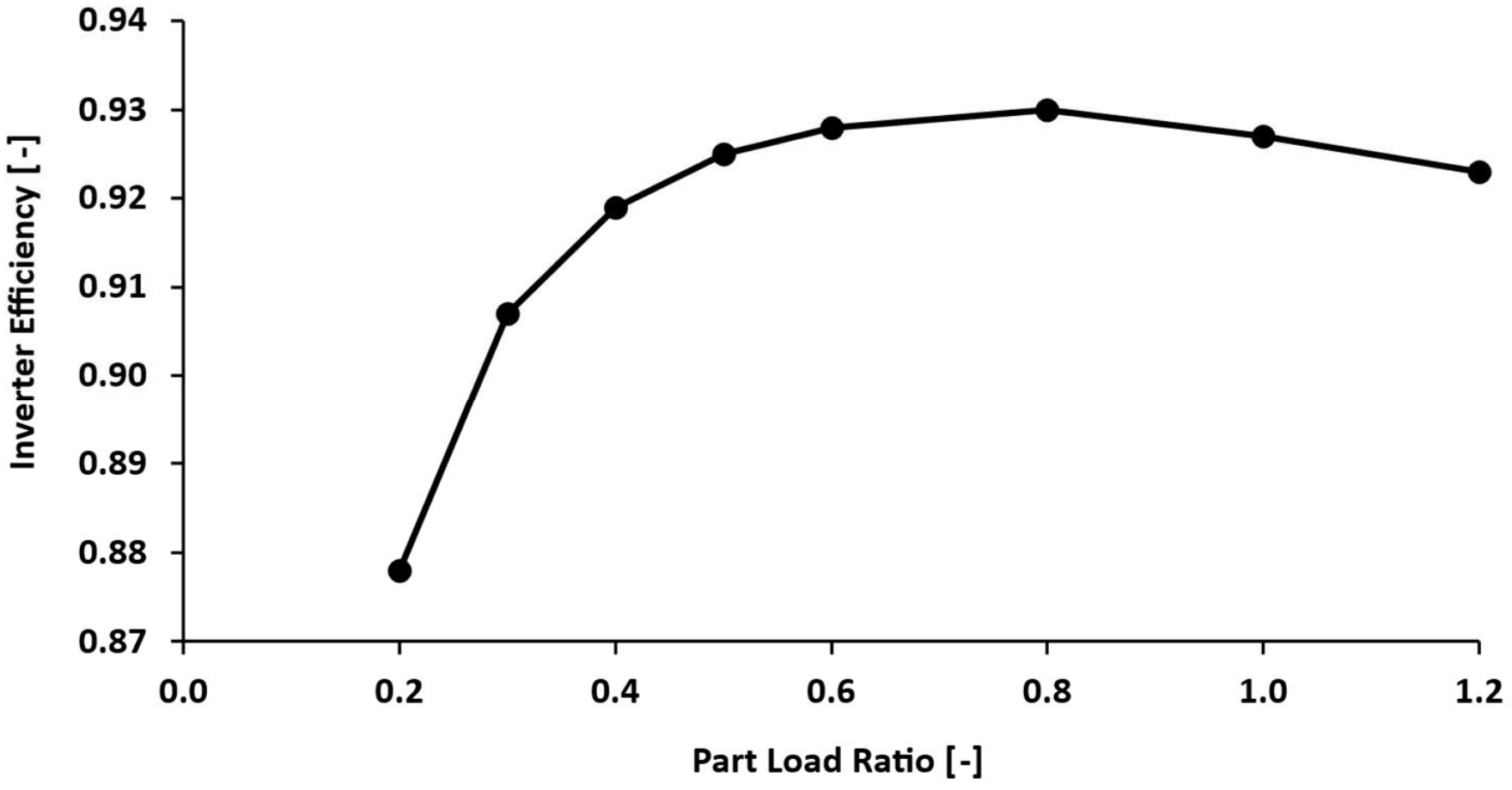

Figure A3: Inverter efficiency curve under different part-load operating conditions.

Table A2: Assumed technical, optical, thermal, and electrical parameters used for the photovoltaic system performance model.

| Name | Symbol | Value | Unit |
|---|---|---|---|
| **PV Module Width** | $W_{PV}$ | 1,134 | mm |
| **PV Module Height** | $H_{PV}$ | 1,762 | mm |
| **PV Module Power** | $P_{module}$ | 440 | W |
| **Angular Losses Coefficient** | $a_r$ | 0.1690 | [-] |
| **Fitting Parameter 1** | $c_1$ | 0.4244 | [-] |
| **Fitting Parameter 2** | $c_2$ | -0.069 | [-] |
| **Mounting Type Correction Factor** | $\omega$ | 1.8 | [-] |
| **PV Thermal Coefficient** | $\beta$ | -0.0035 | $°C^{-1}$ |
| **PV Reference Efficiency** | $\eta_{PV}$ | 0.22 | [-] |
| **PV Balance of System Efficiency** | $\eta_{BOS}$ | 0.904 | [-] |
| **PV Reference Temperature** | $T_{ref}$ | 25 | °C |
| **PV tilt** | $\alpha_{PV}$ | 25 | ° |
| **Inverter Nominal Size** | $P_{inv}$ | 10.0 | kW |

**7.6 Economic Assumptions**

The economic assumptions used in this study are summarized in Table A3. The adopted values were derived from literature data and commercial information from optical fiber and daylighting-system component suppliers. Literature sources were used for the reference costs of conventional vertical farming equipment, operation and maintenance costs, PV-related parameters, and lettuce selling price. Commercial data were used to estimate the costs of the Fresnel lens, solar tracking unit, mechanical frame, UV-IR filter, and the different optical fiber types considered in the analysis.

The capital expenditure of the PV system was estimated starting from the statistical cost data reported in [43] for 8 kW photovoltaic systems. These data were used to recalibrate the cost correlation proposed by the National Institute of Standards and Technology (NIST) [44]. This correlation expresses the total PV system cost as the sum of a fixed cost component and a linear size-dependent component, with the latter expressed in $ $W^{-1}$. The recalibrated correlation adopted in this study is reported in Eq. (A34). When specific references were not available, conservative assumptions were introduced to provide a preliminary estimate consistent with the techno-economic scope of the study.

$$CAPEX_{PV} = 1{,}505 + 2.5 \cdot P_{PV} \quad \text{(A34)}$$

Table A3: Economic Assumptions

| Name | Assumed Cost | Unit | Reference |
|---|---|---|---|
| **Lighting System** | 3.75 | $ $W^{-1}$ | [13] |
| **HVAC System** | 0.65 | $ $W^{-1}$ | [13] |
| **Fresnel Lens** | 380 | $ $m^{-2}$ | [24] |
| **Sun Tracking System** | 175 | $ $unit^{-1}$ | [24] |
| **Mechanical Frame and Mechanical Structure** | 100 | $ $unit^{-1}$ | [51] |
| **UV-IR Filter** | 104 | $ $bundle^{-1}$ | [24] |
| **POF (D = 1 mm, NA = 0.32)** | 2.60 | $ $m^{-1}$ | [40] |
| **POF (D = 1 mm, NA = 0.50)** | 0.86 | $ $m^{-1}$ | [39] |
| **POF (D = 3 mm, NA = 0.50)** | 2.86 | $ $m^{-1}$ | [41] |
| **SOF (D = 1 mm, NA = 0.37)** | 33.0 | $ $m^{-1}$ | [42] |
| **SOF (D = 2 mm, NA = 0.37)** | 50.0 | $ $m^{-1}$ | [-] |
| **Daylighting O&M** | 2% of CAPEX | $ $year^{-1}$ | [-] |
| **Tilted Frame for PV System** | 57.3 | $ $m^{-2}$ | [45] |
| **PV O&M** | 30.36 | $ $kW^{-1}$ | [43] |
| **Lettuce Selling Cost** | 7.82 | $ $kg^{-1}$ | [7] |

### 7.7 Technical Parameters for MIN strategy

Table A4: Geometrical and layout parameters of the Fresnel concentrator and optical fiber receiver for each investigated fiber typology for MIN strategy.

| Parameter | POF1 | POF2 | POF3 | SOF1 | SOF2 | Unit |
|---|---|---|---|---|---|---|
| Focal Length | 800.0 | 450.0 | 450.0 | 650.0 | 650.0 | mm |
| Axial Displacement | 15.0 | 10.0 | 10.0 | 0.0 | 0.0 | mm |
| Receiver Width | 20.0 | 24.0 | 24.0 | 10.0 | 10.0 | mm |
| Collector Spacing | 850.0 | 500.0 | 500.0 | 700.0 | 700.0 | mm |
| Collector Length | 5,000 | 4,000 | 4,000 | 2,250 | 2,250 | mm |
| Collector Number | 9 | 14 | 14 | 10 | 10 | Unit |
| Bundle Number | 2,250 | 2,250 | 2,250 | 2,250 | 2,250 | Unit |
| OF Number | 1,037,250 | 1,496,250 | 164,250 | 258,750 | 63,000 | Unit |

### 7.8 Tonatiuh Ray-Tracing Setup

Ray-tracing simulations were performed using Tonatiuh. The solar source was represented using a pillbox sunshape with a maximum angular radius of 4.65 mrad. The Fresnel lens was modeled as a non-dispersive refractive material with a constant refractive index of 1.494, while the surrounding medium was assumed to be air with n equal to 1. Therefore, wavelength-dependent variations of the refractive index across the PAR range were not explicitly resolved in the ray-tracing simulations. Surface optical imperfections were modeled through a normal slope-error distribution with $\sigma_{slope}$ equal to 2, while no separate specularity error was imposed. A ray-count sensitivity analysis was performed by progressively increasing the number of traced rays; no appreciable changes in the receiver flux distribution were observed beyond 1,000,000 rays, which was therefore adopted for the presented simulations.

**Data Availability Statement:**

The weather data, Tonatiuh scene files, and supporting simulation input files used in this study are available from the corresponding author upon reasonable request. The underlying transient simulation code is part of proprietary licensed software and cannot be distributed.

**Declaration of generative AI and AI use**

During the preparation of this work, the authors used ChatGPT 4 in order to improve the readability of the manuscript. After using this tool/service, the authors reviewed and edited the content as needed and take full responsibility for the content of the published article.